\documentclass[submission,copyright,creativecommons]{eptcs}
\providecommand{\event}{MCU 2013} 
\usepackage{breakurl}             
\usepackage{graphicx}
\def\ligne#1{\hbox to \hsize {#1}}
\newtheorem{tab}{{\sc Table}}
\newtheorem{thm}{{\sc Theorem}}
\newtheorem{lem}{{\sc Lemma}}
\newtheorem{cor}{{\sc Corollary}}
\newtheorem{fig}{{\sc Figure}}
\def\leurre{\noindent\leftskip0pt\footnotesize\baselineskip 10pt\parindent0pt}
\def\grostrait{\ligne{\vrule height 1pt depth 1pt width \hsize}}
\def\demitrait{\ligne{\vrule height 0.5pt depth 0.5pt width \hsize}}
\def\qed{\hfill%
\hbox{
\vbox to 7pt{\hsize=7pt
\hrule height0pt depth0.6pt width\hsize
\hbox to 7pt{\vrule height7pt depth0pt width0.6pt
             \hfill
             \vrule height7pt depth0pt width0.6pt}
\hrule height0.6pt depth0pt width\hsize
}}
}
\def\zz{\tt\footnotesize}
\title{About Strongly Universal Cellular Automata}
\author{Maurice {\sc Margenstern}, 
\institute{Universit\'e de Lorraine,\\ LITA EA3097,\\ Campus du Saulcy,\\
57045 C\'edex, {\sc France}
}
\email{maurice.margenstern@univ-lorraine.fr,{\tt $\backslash$margenstern@gmail.com}}
}
\def\titlerunning{About Universal Cellular Automata}
\def\authorrunning{M. Margenstern}
\begin{document}
\maketitle

\begin{abstract}
In this paper, we construct a strongly universal cellular automaton on the line with
$11$~states and the standard neighbourhood. We embed this construction into several 
tilings of the hyperbolic plane
and of the hyperbolic $3D$ space giving rise to strongly universal cellular automata 
with $10$~states. 
\end{abstract}

\section{Introduction}
\label{intro}

    Many papers about universality of cellular automata, especially those which try to
minimize the number of states preserving this property, consider what is called 
{\bf weakly universal} cellular automata. This term means that the initial configuration
of the cellular automaton may be infinite, provided the following requirement is satisfied:
the initial configuration must be periodic outside a bounded domain. In the case of the 
one dimensional line, it is accepted that the periodicity in the left-hand side infinite part
may be different from the periodicity of the right-hand side one.

   In this paper, we consider deterministic cellular automata in various settings.
First, on the line with the {\bf standard neighbourhood} which means, for each cell, the 
cell itself and its left- and right-hand side neighbours. Second, we shall implement the 
results found on the line into deterministic cellular automata in several tilings of
the hyperbolic plane and one tiling of the hyperbolic $3D$-space.

 For these automata, we consider the construction 
of a {\bf strongly} universal cellular automaton. There are two constraints in order to 
obey this definition which we take from Codd's own definition, see~\cite{codd}. The 
first restriction means that the initial configuration of the cellular automaton must be 
finite. This means that all cells except finitely many of them must be in a quiescent 
state at the initial state. We remind the reader that a state~$s$ is quiescent if and only
if a cell is in~$s$ and its neighbours are also all in~$s$ then it remains in~$s$ at the 
next time. We shall also refer to this situation as {\bf starting from a blank background}.
The second restriction is about the halting: the computation of the cellular
automaton halts if and only if there is the occurrence of two consecutive identical
configurations. Now, it is in the definition of universality that when the simulated
device halts, the simulating device has also to stop. This is the counterpart of the situation
when the simulated device does not halt: then, by the very notion itself, the simulating
device cannot halt. Now, few papers take into account this halting condition, although it is
a natural one. Most papers follow different definitions of universality. Probably, the most 
studied case is what is called {\bf weak universality} by this author and also a few ones, 
see~\cite{woodsneary} for further references. 
In weak universality, both above constraints are 
removed. For the halting, it is enough that the simulated machine halts, the simulating one
may not halt.
For the initial configuration, infinite ones are accepted provided they are not 
arbitrary. 
The usual constraint is, in the case of a $1D$-cellular automaton that the configuration
be periodic in both directions, starting from a certain point. It is not required that
the periodic pattern should be the same on both sides.
Accordingly much striking results were obtained. Among them, the most notable one is
the cellular automaton with Rule~110, see~\cite{cook,wolfram}. In this case we shall also
say that the cellular automaton works on a non-blank periodic background.

\ifnum 1=0 {
In this paper, we follow the method used in~\cite{lindgren-nordahl} where they
claim that they construct a weakly universal cellular automaton with 7~states only. 
For this purpose, they simulate Minsky's Turing machine with 7~states and 4~letters,
see~\cite{minsky}.
As already pointed at in~\cite{rogozhinMI,rogozhinTCS}, Minsky's machine is not
universal as it damages its tape before halting when the simulated machine halts.
I turn to that point in Section~\ref{strong} in order to explain what happens.
Accordingly, the best which is proved in~\cite{lindgren-nordahl} is a cellular
automaton whose halting is undecidable, although what is meant there by halting is not
that clear. Accordingly, the claim by the authors that
they obtain a universal cellular automaton from a blank background with~9 states is not
grounded. Moreover, in~\cite{lindgren-nordahl}, they do not explicitly indicate how 
to go from~7 states to 9~states in order to start from a blank background.
In this paper, we construct cellular
automaton with 13~states, which is strongly universal in the above meaning: it starts from
a finite configuration and when the simulated computation halts, the cellular automaton
also halts. 
It is possible that the present strongly universal cellular automaton has the smallest
number of states, but I am not sure of that.
There is a strongly universal cellular automaton with 18~states in~\cite{alvy_smith_iii}
and another one with 14~states in~\cite{albert_culik_ii}.
} \fi

   The paper considers the cellular automaton with 7~states constructed 
in~\cite{lindgren-nordahl} which we later refer to as~$\cal LN$. This cellular automaton 
is weakly universal. In~\cite{lindgren-nordahl}, they claim that with two additional states, 
they obtain from~$\cal LN$ a strongly universal cellular automaton. However, they do not give 
any hint on how these additional states are used and, moreover, they do not consider the 
halting of the their cellular automaton as mentioned above. Accordingly, such a result, if 
any, cannot be taken into consideration. Our construction starts from~$\cal LN$ to which
we append four states in order to be strongly universal as stated in the above definition.
Then we transport the construction into two tessellations of the hyperbolic plane and
one of the hyperbolic $3D$~space.

In Section~\ref{implement}, we remind the construction 
of~\cite{lindgren-nordahl} for self-containedness. 
Also in this section, we clarify the status of $\cal LN$ with respect to weak universality.
Indeed, it was suspected to be not possible, due
to a confusion about the Turing machine simulated by~$\cal LN$.
In Section~\ref{strong}, we show how the additional states allow us to transform~$\cal LN$
into a strongly universal cellular automaton with 11~states.

%
%
In Section~\ref{extent}, we extend the result 
to cellular automata in hyperbolic spaces. There, it is possible to obtain strongly universal
cellular automata with~10 states only in the tilings of the spaces which we shall
consider.

\section{Implementing a Turing machine in a one-dimensional cellular automaton}
\label{implement}
  
   Weak universality results for cellular automata are rather easy to obtain 
from the simulation of a Turing machine. The idea is to embed the Turing tape
into the cellular automaton by regularly putting the symbols of the Turing machine,
two neighbouring symbols of the tape being separated by the same number of blanks
of the cellular automaton. 

   In order to obtain a strongly universal cellular automaton, only a finite part of
the Turing tape can be embedded in such a way. In order to perform the computation,
especially if the computation turns out to be infinite, we have also to implement the 
continuation of the Turing tape. For the present moment,
let us ignore this point to which we go back in Section~\ref{strong}. In 
Sub-section~\ref{guideline} we remind the general scheme of simulation, summarizing
the presentation of~\cite{lindgren-nordahl}. 
In Sub-section~\ref{exec}, we explicitly indicate
the Turing machine we consider, showing that it is strongly universal, and how it 
is implemented as in~\cite{lindgren-nordahl}. 

\subsection{The guideline for the simulation}
\label{guideline}
\def\LM{\hbox{$\mathcal L$$(\cal M$$)$}}

Consider a Turing machine~$\cal M$ whose blank symbol is denoted by~{\tt 0}. Later, we shall
explicitly indicate which Turing machine will play the role of~$\cal M$.
Now, for each such machine
$\cal M$, we construct a deterministic cellular automaton which simulates~$\cal M$ 
and we denote this automaton by\LM.

   Let us first indicate how \LM{} works, following the technique given 
in~\cite{lindgren-nordahl}. We say that a cell is {\bf blank} if it
is in the quiescent state. Accordingly, if a cell is blank as well as its left- and 
right-hand side neighbours, it remains blank at the next step of the computation.

\def\theblank{\hbox{$\_\!\_$}}

   The Turing tape is implemented by inserting the squares of the tape at regular places
of the cellular automaton, leaving the same amount of blank cells in between two consecutive
places. This means that outside a finite interval of the support of the cellular automaton,
the pattern \hbox{\tt 0 \theblank$^k$}, where $k$ is the number of blank cells of~\LM{}
in between two symbols of~$\cal M$. In~\cite{lindgren-nordahl}, $k$~is fixed as at least~2.
Here, we also fix \hbox{$k=2$}, as the value \hbox{$k=2$} will play an important role
in Section~\ref{strong}. Accordingly, the initial configuration of~\LM{} 
looks like this: \hbox{$(\theblank$ \tt x $\theblank)^\infty$}

   However, in this representation, we do not know where is the head of the Turing machine
and in which state it is. The Turing head is represented by~{\tt T}, following the
notation of~\cite{lindgren-nordahl}, assuming that {\tt T} is not a letter of~$\cal M$. 
In order to indicate the state, {\tt T} is accompanied
by an auxiliary symbol which represents the state. But this representation is not direct.
As in many papers, the basic idea of~\cite{lindgren-nordahl} is to mimic the space-time 
diagram of the Turing machine by that of~\LM. Under this consideration, {\tt T} appears
as a moving symbol which changes its neighbour: the left-hand side or the right-hand side one,
depending on the direction of the move defined by the current instruction. The idea
of~\cite{lindgren-nordahl} is to append a symbol to the left of~{\tt T} for the
instructions moving to right and to append a symbol to the right of~{\tt T}
for the instructions moving to left. Following~\cite{lindgren-nordahl}, we call
a pair {\tt T a} or {\tt a T} an {\bf impulsion}, to left or to right respectively.
As the impulsions travel on the support of~\LM, they meet with symbols of~{\cal M}.
Such a meeting is called a {\bf collision} in~\cite{lindgren-nordahl} and it is desirable to
organize it in such a way that after the collision the support shows us the next configuration
of the Turing machine: the result of the collision will be the application of the
corresponding instruction.

Below, Tables~\ref{simple} and~\ref{demi-tours} show us how the impulsions travel in the
support of~\LM, and how instructions are performed. Note that in these tables the blank
of~\LM{} is denoted by the character~`.'. The same convention will be used for tables based
on executions by a computer program.

\newdimen\largesitu\largesitu=177pt

\setbox110=\vtop{
\vtop{\leftskip 0pt\parindent 0pt\hsize=\largesitu
\zz
\obeylines
\obeyspaces\global\let =\ \parskip=-2pt
 . u . a T . v . . x . . y . .
\par}
\vskip-8pt
\vtop{\leftskip 0pt\parindent 0pt\hsize=\largesitu
\zz
\ligne{\hrulefill}
\vskip-9pt
\ligne{\hrulefill}
}
\vskip-2pt
\vtop{\leftskip 0pt\parindent 0pt\hsize=\largesitu
\zz
\obeylines
\obeyspaces\global\let =\ \parskip=-2pt
 . u . . w . b T . x . . y . .
\par}
}

\setbox112\vtop{
\vtop{\leftskip 0pt\parindent 0pt\hsize=\largesitu
\zz
\obeylines
\obeyspaces\global\let =\ \parskip=-2pt
 . u . . v . T a . x . . y . .
\par}
\vskip-8pt
\vtop{\leftskip 0pt\parindent 0pt\hsize=\largesitu
\zz
\ligne{\hrulefill}
\vskip-9pt
\ligne{\hrulefill}
}
\vskip-2pt
\vtop{\leftskip 0pt\parindent 0pt\hsize=\largesitu
\zz
\obeylines
\obeyspaces\global\let =\ \parskip=-2pt
 . u . T b . w . . x . . y . .
\par}
}
\vtop{
\begin{tab}\label{simple}
\leurre
Standard situation for an impulsion whose direction is not changed
by the collision. Left-hand side: impulsion to right; right-hand side: impulsion
to left.
\end{tab}
\ligne{\hfill
\box110\hfill\box112\hfill}
\vskip 10pt
}

   Table~\ref{simple} shows the execution of an instruction in the case when the application
of the instruction does not change the direction of the motion of the Turing head. By contrast,
Table~\ref{demi-tours} shows the situations when the execution of the instruction results
in a change in the direction of the motion of the Turing head. This situation is usually
called a {\bf half-turn} of the Turing head. 

\setbox110=\vtop{
\vtop{\leftskip 0pt\parindent 0pt\hsize=\largesitu
\zz
\obeylines
\obeyspaces\global\let =\ \parskip=-2pt
 . u . a T . v . . x . . y . .
\par}
\vskip-8pt
\vtop{\leftskip 0pt\parindent 0pt\hsize=\largesitu
\zz
\ligne{\hrulefill}
\vskip-9pt
\ligne{\hrulefill}
}
\vskip-2pt
\vtop{\leftskip 0pt\parindent 0pt\hsize=\largesitu
\zz
\obeylines
\obeyspaces\global\let =\ \parskip=-2pt
 . u . T b . w . . x . . y . .
\par}
}

\setbox112\vtop{
\vtop{\leftskip 0pt\parindent 0pt\hsize=\largesitu
\zz
\obeylines
\obeyspaces\global\let =\ \parskip=-2pt
 . u . . v . T a . x . . y . .
\par}
\vskip-8pt
\vtop{\leftskip 0pt\parindent 0pt\hsize=\largesitu
\zz
\ligne{\hrulefill}
\vskip-9pt
\ligne{\hrulefill}
}
\vskip-2pt
\vtop{\leftskip 0pt\parindent 0pt\hsize=\largesitu
\zz
\obeylines
\obeyspaces\global\let =\ \parskip=-2pt
 . u . . w . b T . x . . y . .
\par}
}
\vtop{
\begin{tab}\label{demi-tours}
\leurre
Situations of a standard half-turn. Left-hand side: half-turn to left; right-hand side: half-turn
to right.
\end{tab}
\ligne{\hfill
\box110\hfill\box112\hfill}
\vskip 10pt
}

    In Tables~\ref{simple} and~\ref{demi-tours}, we can say that the upper line represents
the {\bf starting point} of the collision and that the lower line, below the double bars,
represents the {\bf final point}. We can notice that the starting point represents a 
configuration of~$\cal M$ and that the final one represents the next configuration of~$\cal M$.
However, the final point is never the next configuration of the starting point for~\LM.
As can be noticed on the figures, the final point occurs at least three steps after the
starting point. In~\cite{lindgren-nordahl}, the number of steps of~\LM{} between the starting 
point and the final one can be higher, and this will also be the case in our simulation.

Say that a configuration by the end of a collision is a {\bf standard} one
if it is of the form \hbox{\tt z . \hskip-7pt y T} or \hbox{\tt T y . \hskip-7pt z}, where {\tt z} is
the new state after the collision which corresponds to the execution of the current
instruction of the Turing machine on the currently scanned symbol.
From Tables~\ref{simple} and~\ref{demi-tours}, we can see that if we consider three 
consecutive Turing symbols
where the middle one is scanned by an impulsion then, after the collision is
completed, the positions of the extremal Turing symbols are unchanged. This ensures
that the working of the cellular automaton faithfully simulates the Turing machine.

Now, there are a few exceptional situations illustrated by Table~\ref{double},
where the new state of the scanned symbol appears one step later with respect 
to a standard configuration. 

However, if we look at the extension in space, we may have the situation illustrated by
Table~\ref{double}, but no more. We shall check them in Section~\ref{strong}. This indeed
occurs in~\cite{lindgren-nordahl} with a few collisions. 

\setbox110=\vtop{
\vtop{\leftskip 0pt\parindent 0pt\hsize=\largesitu
\zz
\obeylines
\obeyspaces\global\let =\ \parskip=-2pt
 . u . a T . v . . x . . y . .
\par}
\vskip-8pt
\vtop{\leftskip 0pt\parindent 0pt\hsize=\largesitu
\zz
\ligne{\hrulefill}
\vskip-9pt
\ligne{\hrulefill}
}
\vskip-2pt
\vtop{\leftskip 0pt\parindent 0pt\hsize=\largesitu
\zz
\obeylines
\obeyspaces\global\let =\ \parskip=-2pt
 . u . . w . b T . x . . y . .
 . u . . z . . b c x . . y . 
\par}
}

\setbox112\vtop{
\vtop{\leftskip 0pt\parindent 0pt\hsize=\largesitu
\zz
\obeylines
\obeyspaces\global\let =\ \parskip=-2pt
 . u . . v . T a . x . . y . .
\par}
\vskip-8pt
\vtop{\leftskip 0pt\parindent 0pt\hsize=\largesitu
\zz
\ligne{\hrulefill}
\vskip-9pt
\ligne{\hrulefill}
}
\vskip-2pt
\vtop{\leftskip 0pt\parindent 0pt\hsize=\largesitu
\zz
\obeylines
\obeyspaces\global\let =\ \parskip=-2pt
 . u . . z . b T . x . . y . .
 . u . . w . . b c x . . y . .
\par}
}
\vtop{
\begin{tab}\label{double}
\leurre
In these cases, there is delay in the transformation of the new state by one step.
Left-hand side: for an impulsion to right not changed by the collision;
right-hand side: for a half-turn to right. There are no other cases involving such a delay. 
\end{tab}
\ligne{\hfill
\box110\hfill\box112\hfill}
\vskip 10pt
}

   Let us now look at how many symbols are needed. The impulsions correspond to the
indication of a state and of a move. If we analyze the table of~$\cal M$ in this regard,
there are $s_\ell$~states in which all instruction go to the left, $s_r$~of them, where
all instructions go to the right and $s_b$~states where we meet instructions of both kinds. 
Of course,
\hbox{$s_\ell$+$s_r$+$s_b=n$}, where $n$~is the number of states of~$\cal M$.
Let us set \hbox{$n_\ell=s_\ell$+$s_b$} and \hbox{$n_r=s_r$+$s_b$}. Then,
the coding of the instructions to the left will require $n_\ell$~symbols while the coding
of those to the right will require~$n_r$~of them. In any case, we need at least
\hbox{$\mu=\hbox{\rm max}(n_\ell,n_r)$} symbols to code the number of impulsions. The same symbol
may be used to code one impulsion to the left and one impulsion to the right. This
is why in Tables~\ref{simple}, \ref{demi-tours} and~\ref{double}, we cannot interpret the
symbol which is appended to~{\tt T} as a state~: it is simply an encoding of the impulsion.
If \hbox{$\mu\leq m$}, where $m$~is the number of letters of~$\cal M$, then we need
just two extra symbols: {\tt T} and \theblank, the blank of~\LM. If \hbox{$m<\mu$}
we still need \hbox{$\mu$$-$$m$} symbols. Note that the number of
collisions is \hbox{$m(n_\ell$+$n_r)$}.

In the case of~\cite{lindgren-nordahl},
$\mu=5$ and $m=4$. This is why we need at least 7~symbols. It turns out that, using
an infinite background as indicated above, with the Turing machine used
in~\cite{lindgren-nordahl}, it is enough.

\newdimen\lalarge\lalarge=75pt
\newdimen\largeura\largeura=50pt
\newdimen\largeur\largeur=40pt
\def\lignetab #1 #2 #3 #4 #5 {%
\hbox{\hbox to \largeura{\hfill#1\hfill}
\hbox to \largeur{\hfill#2\hfill}
\hbox to \largeur{\hfill#3\hfill}
\hbox to \largeur{\hfill#4\hfill}
\hbox to \largeur{\hfill#5\hfill}
}\vskip 6pt}

\newdimen\largeLR\largeLR=53.1pt
\def\macrostepLR #1 #2 #3 #4 {%
\vtop{\leftskip 0pt\parindent 0pt\hsize=\largeLR
\ligne{\zz \ \ #1\ \theblank\ T\ #2\hfill} 
\vskip-9pt
\ligne{\hrulefill}
\vskip-5pt
\vskip-7pt
\ligne{\hrulefill}
\vskip-3pt
\ligne{\zz \ \ #3\ \theblank\ #4\ T\hfill} 
}} 

\newdimen\largeLL\largeLL=53.1pt
\def\macrostepLL #1 #2 #3 #4 {%
\vtop{\leftskip 0pt\parindent 0pt\hsize=\largeLL
\ligne{\zz \ \ \ #1\ \theblank\ T\ #2\hfill} 
\vskip-9pt
\ligne{\hrulefill}
\vskip-5pt
\vskip-7pt
\ligne{\hrulefill}
\vskip-3pt
\ligne{\zz \ T\ #3\ \theblank\ #4\hfill} 
}} 

\newdimen\largeRL\largeRL=53.1pt
\def\macrostepRL #1 #2 #3 #4 {%
\vtop{\leftskip 0pt\parindent 0pt\hsize=\largeRL
\ligne{\zz \ \ #1\ T\ \theblank\ #2\hfill} 
\vskip-9pt
\ligne{\hrulefill}
\vskip-5pt
\vskip-7pt
\ligne{\hrulefill}
\vskip-3pt
\ligne{\zz \ \ T\ #3\ \theblank\ #4\hfill} 
}} 

\newdimen\largeRR\largeRR=53.1pt
\def\macrostepRR #1 #2 #3 #4 {%
\vtop{\leftskip 0pt\parindent 0pt\hsize=\largeRR
\ligne{\zz \ #1\ T\ \theblank\ #2\hfill} 
\vskip-9pt
\ligne{\hrulefill}
\vskip-5pt
\vskip-7pt
\ligne{\hrulefill}
\vskip-3pt
\ligne{\zz \ \ \ #3\ \theblank\ #4\ T\hfill} 
}} 

\newdimen\largeLS\largeLS=50pt
\def\macrostepLS #1 #2 #3 {%
\vtop{\leftskip 0pt\parindent 0pt\hsize=\largeLS
\ligne{\zz \ \ #1\ \theblank\ T\ #2\hfill} 
\vskip-9pt
\ligne{\hrulefill}
\vskip-5pt
\vskip-7pt
\ligne{\hrulefill}
\vskip-3pt
\ligne{\zz \ \ \theblank\ #3\ \theblank\hfill} 
}} 

\newdimen\largeRS\largeRS=50pt
\def\macrostepRS #1 #2 #3 {%
\vtop{\leftskip 0pt\parindent 0pt\hsize=\largeRS
\ligne{\zz #1\ \theblank\ #2\ T} 
\vskip-9pt
\ligne{\hrulefill}
\vskip-5pt
\vskip-7pt
\ligne{\hrulefill}
\vskip-3pt
\ligne{\zz \theblank\ #3\ \theblank} 
}} 

\subsection{Implementing Minsky's Turing machine}
\label{exec}

   Let us now turn to the Turing machine used in~\cite{lindgren-nordahl}.
In Sub-subsection~\ref{minskystrong}, we explain why Minsky's Turing machine is
strongly universal. In Sub-subsection~\ref{trueexec}, we remind its implementation
in~\cite{lindgren-nordahl}.

\subsubsection{A strongly universal Turing machine}
\label{minskystrong}
 
This machine was devised by M.~Minsky in~\cite{minsky}: it has seven states and four letters
and, in~\cite{minsky}, it was claimed to be strongly universal. We say {\it claimed} because
the proof given in~\cite{minsky} cannot considered as completely satisfactory. This is
why in~\cite{rogozhinTCS,rogozhinMI,robinsonMT,mmstrong}, several authors, including myself, 
considered that this machine is not universal: it was said that it damages its
output before halting when the simulated machine halts. In~\cite{woodsneary}, the authors
claim that Minsky's machine is indeed universal, but they do not give enough details to
understand why it is the case. After discussing this point with Turlough Neary, I came
to the conclusion that the claim of~\cite{woodsneary} is correct. After sometime, I also
understood where the reason is exactly. I think it useful to clarify this point, which
I try to do in some general form which seems to be interesting in itself.

   Table~\ref{minskyUTM} gives the table of Minsky's machine which we later refer to
as Minsky's UTM. We follow the following 
notations, inspired by those of~\cite{minsky}. An instruction of a Turing machine
is of the form \hbox{\tt p$xMy$q}. The head reads~$x$ in state~{\tt p}. It then
replaces~$x$ by~$y$, moves to the left-, right-hand side neighbour or remains on the same
cell depending on whether $M$~is $L$, $R$~or~$S$ and the heads is now under the state~{\tt q}. 
We read~$x$ in the heads of the columns and~{\tt p} on the heads of the lines in the Table. The 
corresponding entry contains the triple \hbox{$My$\tt q} of the instruction. In the table, we 
always define~$M$  and we mention $y$ or~{\tt q} only if it is different from~$x$, {\tt p}
respectively.

   As proved in~\cite{coke-minsky,minsky}, this machine is universal as it is able to simulate
a tag-system with deletion number~2 which was proved to be universal in the same references.
A tag-system itself is a particular kind of Post system. It consists of a finite alphabet~$A$ 
together with a mapping~$p$ from~$A$ into~$A^\star$ the set of words on~$A$. The tag-system works
as follows. First, it reads the first two letters of the current tagged word~$w$, 
say $\ell_1\ell_2$.
Then, it erases $\ell_1\ell_2$ from the beginning of~$w$. This is why we say that
this tag-system has deletion number~2. Then, the tag-system appends to what
remains of~$w$ the word $p(\ell_1)$: this is the new configuration of the tagged word after
the application of one step of the tag-system. Among the letters of~$A$, one of them at least 
is a~{\bf halting letter}: when the tag-system meets such a letter as the first letter of the
current word, it performs the step and when this is completed, it halts the computation.
When it halts, the computation of the tag system can be described as a sequence of
$w_0$, ..., $w_n$, where $w_0$ is the initial tagged word and
$w_n$ is the last one, subjected to the following condition: $w_{i+1}$ is the word obtained
by the computation of the tag-system by the completion of the step which erased the last 
letter of~$w_i$. We say that the transformation of~$w_i$ into~$w_{i+1}$ defines a
{\bf pass} of the tag-system over~$w_i$. Note that a current tagged word is not necessary
one of the $w_i$'s.
Minsky imposes two assumptions on the
tag-system as just described: there is a single halting letter and the tagged word must always 
be non-empty during the computation. It is an easy 
exercise, left to the reader, to check that these conditions are observed by the
the tag system devised by Cocke and Minsky in their proof. The full solution is already 
in~\cite{minsky}. 

\vtop{
\begin{tab}\label{minskyUTM}
\leurre
Table of Minsky's Turing machine with $7$~states and $4$~letters, from~{\rm\cite{minsky}}.
\end{tab}
\vskip 5pt
\ligne{\hfill
\vtop{\leftskip 0pt\parindent 0pt\offinterlineskip\hsize=250pt
\lignetab {} 0 1 {$y$} {$A$}
\vskip 5pt
\lignetab {1} {$L$} {$L2$} {$0L$} {$1L$}
\lignetab {2} {$yR$} {$AR$} {$0L1$} {$yR6$}
\lignetab {3} {$\_\!\!\_\!\!\_\!\!\_$} {$AL$} {$L$} {$1L$4} 
\lignetab {4} {$yR5$} {$L7$} {$L$} {$1L$}
\lignetab {5} {$yL3$} {$AR$} {$R$} {$1R$}
\lignetab {6} {$AL3$} {$AR$} {$R$} {$1R$}
\lignetab {7} {$yR6$} {$R$} {$0R$} {$0R2$}
}
\hfill}
\vskip 5pt
}

   It is interesting to understand why some people claimed that Minsky's UTM damages
its tape and why this is not necessary the case. Here I explain where the reason is.

   Define an {\bf extended tag-system} with deletion number~2 as given by two 
finite alphabets, $A$ and~$T$, together with a mapping~$p$ from~$A$ into words 
on~$(A+T)^\star$. We call~$T$ the terminal alphabet. The extended tag-system works 
exactly as a tag-system with deletion number~2 and we have exactly the same notion of pass. 
However, a halting computation is considered 
as valid only when a terminal letter is never the first letter of the current word during the
computation, except possibly the final one, and the last current word is in $T^\star$. 
A usual tag-system is of course an extended one whose terminal alphabet is the
whole alphabet.
We have the following property which, I think, is interesting {\it per se}:

\begin{lem}\label{extended} $-$
Let $E$ be an extended tag-system with deletion number~$2$. There is a tag-system~$P$
also with deletion number~2, which simulates~$E$ in the following meaning:
if $E$ does not halt on~$w$, $P$ also does not halt. If \hbox{$w=w_1..w_n$} is accepted 
by~$E$, then $P$ halts on~$\overline{w}$ if $n$~is odd, on~$a\overline{w}$ if $n$~is
even,
with \hbox{$\overline{w} = s_1s_1..s_ks_k$} where $s_i$ are defined by 
\hbox{$s_i=w_{2i+1}$}, \hbox{$i\in\{1..\lfloor \displaystyle{{n-1}\over2}\rfloor\}$}
and $a\in T$.
\end{lem}

\noindent
Proof. Let $A$ and~$T$ be the alphabets of~$E$, $T$ being the terminal one. Let $p$~be the
mapping from~$A$ into~$(A+T)^\star$ defining the computation of~$E$. We construct~$P$
as follows. Its alphabet is~$(A+T+\{z\})^\star$. Define the mapping $q$ of~$P$ 
by~\hbox{$q(a)=p(a)$} for $a\in A\backslash T$,
$aT$ not being the halting letter. For the halting letter~$h$, \hbox{$q(h)=p(h)zza$}, $a$
being any letter of~$T^\star$,
with the addition that~$h$ is
not a halting letter for~$P$. For $b$~in~$T$,
\hbox{$q(b) = bb$}. The halting letter is~$z$ and \hbox{$q(z)=\epsilon$} where $\epsilon$
is the empty word.

   From the definition of~$P$, it is clear that when the computation of~$E$ on~$w$ does not
halt, the computation of~$P$ on the same word does not halt. If the computation of~$E$
on~$w$ is accepted, it halts and the final word is~$\omega$ with~$\omega\in T^\star$.
The computation with~$P$ replaces $\omega$ by~$\omega zza$ as far as, by definition,
$\omega$ ends with~$p(h)$ which is in~$T^\star$ by our assumption. Now, as all letters 
of~$\omega$ are in~$T$,
$P$ transforms $\omega zza$ into $zza\overline{\omega}$
or $za\overline{\omega}$ depending of the parity of $\vert\omega\vert$ as
mentioned in the statement of the lemma. Accordingly, when $P$ is applied
to $za\overline{\omega}$ or~$zza\overline{\omega}$, it produces $\overline{\omega}$ 
or~$a\overline{\omega}$ and then, it halts.
\qed

   Now, Minsky's UTM does not simulate a tag system~$P$ but, rather, some extension of~$P$. 
Indeed, when Minsky's UTM simulates the halting
step of~$P$, it copies the pattern {\tt AAA} at the end of the configuration and it does not 
stop. The halting of Minsky's UTM requires that at some point, the head in state~3 
meets a~{\tt 0} on the tape. The tape of Minsky's UTM contains an encoding of the productions
on the left-hand side and an encoding of the tagged word on the right-hand side with a possible
garbage of {\tt0}'s in between.
The computation can be divided into cycles, each one of them
being split into three steps: first, the localization of the production corresponding
to the first letter; second, it copies the production. During these two steps, an
appropriate marking of the configuration from the tagged word to the productions
allows the machine to perform these tasks. The third stage consists in removing the 
markings and in detecting the first letter of the current word. The first step is 
accomplished by states~1 and~2. The copying is performed by states~3, 4, 5 and~6. 
The restoration and the transition to the next cycle are performed by states~6 and~7. 
During the copying, state~5 copies~{\tt 0}'s 
and state~6 copies~{\tt 1}'s, going from the productions to the end of the current word.
When the symbol, {\tt 0} or~{\tt 1} is copied, the return of the head to the productions
is performed both by states~3 and~4: state~3 leads the head until it meets an~{\tt A}
at which moment state~4 takes the control, leading the head back to the production. 
It is important to note that this transition from state~3 to state~4 is made possible 
by the markings of
the first steps which marks the configuration by {\tt y}'s and~{\tt A's} from the
beginning of the current word to the last copied symbol in the productions.
State~3 removes the markings in the current word and state~4 removes those of the production,
the~{\tt y}'s being still present from the current word to the productions. State~7
changes back those~{\tt y}'s to~{\tt 0}.
When the halting letter of the tag-system is detected, Minsky's UTM
copies the halting production at the end of the configuration. This consists in
the pattern~{\tt AAA}. But the machine does not stop, so that it starts a new pass of
the tag-system over the final word. As long as codes of words remain on the left-hand side 
of~{\tt AAA}, the computation of the tag system goes on, performing the
last pass in Lemma~\ref{extended}.
Due to the erasing process, the machine eventually meets~{\tt AAA} in state~7.
The leftmost~{\tt A} read under this state signals a new encoded letter. 
But the second~{\tt A} read under state~2 shows that the end of the word is reached.
And so there is no marking: the machine under state~6 appends 
an~{\tt A} at the right-hand side end of the configuration. When it goes back in state~3, 
it will then meet a~{\tt 0} in this state as there is no marking in between {\tt AAA}
and the production. And so, the machine will halt.

    This working reminds us that of an extended tag-system. 
From our lemma, it is plain that if we can present the computation of the execution
of an extended tag-system and if we have $\omega=\overline\omega$ for the final
word~$\omega$, then we can recover the production of the initial tag-system.
It is namely possible for the tag-system~$\cal T$ devised 
by Cocke-Minsky to prove that tag-system with deletion number~2 can simulate any Turing 
machine on~$\{0,1\}$. This tag-system encodes the configuration of a Turing machine with
four letters as follows:

\vskip 5pt
\ligne{\hfill\tt Aa(aa)$^n$Bb(bb)$^m$\hfill}
\vskip 5pt

\noindent
where $n$ and~$m$ are non-negative integers which encode the left- and the right-hand part,
respectively, of the Turing tape, starting from the position of its head.
We could change \hbox{\tt Aa(aa)$^n$Bb(bb)$^m$} to:
\vskip 5pt
\ligne{\hfill \hbox{\tt AA(aa)$^n$BB(bb)$^m$}\hfill}
\vskip 5pt
\noindent
in order to be compatible with  Lemma~\ref{extended}. This does not change the proof
of~\cite{coke-minsky,minsky}.
Now these four letters represent the states of the Turing machine: when an
instruction of the Turing machine is performed, a new group of four letters appear,
representing the new state of the head of the Turing machine. Accordingly, the final
state is associated to a group of four letters~:

\vskip 5pt
\ligne{\hfill\tt Hh(hh)$^n$Kk(kk)$^m$\hfill}
\vskip 5pt
\noindent
where {\tt H} is the halting letter of~$\cal T$. Note that {\tt h}, {\tt K} and {\tt k}
never trigger a production of the tag-system. Note that here too, 
\hbox{\tt Hh(hh)$^n$Kk(kk)$^m$} can be
changed to \hbox{\tt HH(hh)$^n$KK(kk)$^m$} without altering the proof.
Accordingly, the simulation 
of the Turing machine by~$\cal T$ can be presented as the computation in an extended 
tag-system. 
Moreover, this tag system has clearly the property that
$w=\overline w$ for all current word~$w$ during the computation. 
So that the lemma tells us how to transform~$\cal T$ 
into a standard tag-system~$P$ which computes exactly the same thing as~$\cal T$.
From the proof of the lemma and the behaviour of Minsky's UTM as we presented above
show us that Minsky's UTM exactly executes~$P$, which proves the strong universality
of Minsky's UTM.

\subsubsection{The Lindgren-Nordahl cellular automaton with 7 states}
\label{trueexec}

   The implementation of Minsky's UTM in a cellular automaton was performed
in~\cite{lindgren-nordahl} as mentioned in Subsection~\ref{guideline}.

   For the sake of the reader, we mention the exact implementation.

   From Table~\ref{minskyUTM}, we can see that we have the following impulsions
to right: states~2, 5, 6 and~7 and that we have five impulsions to left:
states~1, 2, 3, 4 and 7. They are accordingly encoded as
\hbox{\tt yT$,$ 0T$,$ 1T$,$ AT} and \hbox{\tt Ty$,$ T0$,$ T1$,$ TA$,$ TB}
respectively.

\newdimen\largeLR\largeLR=53.1pt
\def\macrostepLR #1 #2 #3 #4 {%
\vtop{\leftskip 0pt\parindent 0pt\hsize=\largeLR
\ligne{\zz #1\ \theblank\ T\ #2} 
\vskip-9pt
\ligne{\hrulefill}
\vskip-5pt
\vskip-7pt
\ligne{\hrulefill}
\vskip-3pt
\ligne{\zz #3\ \theblank\ #4\ T} 
}} 

\newdimen\largeLL\largeLL=53.1pt
\def\macrostepLL #1 #2 #3 #4 {%
\vtop{\leftskip 0pt\parindent 0pt\hsize=\largeLL
\ligne{\zz \ \ #1\ \theblank\ T\ #2} 
\vskip-9pt
\ligne{\hrulefill}
\vskip-5pt
\vskip-7pt
\ligne{\hrulefill}
\vskip-3pt
\ligne{\zz T\ #3\ \theblank\ #4} 
}} 

\newdimen\largeRL\largeRL=53.1pt
\def\macrostepRL #1 #2 #3 #4 {%
\vtop{\leftskip 0pt\parindent 0pt\hsize=\largeRL
\ligne{\zz #1\ T\ \theblank\ #2} 
\vskip-9pt
\ligne{\hrulefill}
\vskip-5pt
\vskip-7pt
\ligne{\hrulefill}
\vskip-3pt
\ligne{\zz T\ #3\ \theblank\ #4} 
}} 

\newdimen\largeRR\largeRR=53.1pt
\def\macrostepRR #1 #2 #3 #4 {%
\vtop{\leftskip 0pt\parindent 0pt\hsize=\largeRR
\ligne{\zz #1\ T\ \theblank\ #2} 
\vskip-9pt
\ligne{\hrulefill}
\vskip-5pt
\vskip-7pt
\ligne{\hrulefill}
\vskip-3pt
\ligne{\zz \ \ #3\ \theblank\ #4\ T} 
}} 

\newdimen\largeLS\largeLS=50pt
\def\macrostepLS #1 #2 #3 {%
\vtop{\leftskip 0pt\parindent 0pt\hsize=\largeLS
\ligne{\zz #1\ \theblank\ T\ #2} 
\vskip-9pt
\ligne{\hrulefill}
\vskip-5pt
\vskip-7pt
\ligne{\hrulefill}
\vskip-3pt
\ligne{\zz \theblank\ #3\ \theblank} 
}} 

\newdimen\largeRS\largeRS=50pt
\def\macrostepRS #1 #2 #3 {%
\vtop{\leftskip 0pt\parindent 0pt\hsize=\largeRS
\ligne{\zz #1\ \theblank\ #2\ T} 
\vskip-9pt
\ligne{\hrulefill}
\vskip-5pt
\vskip-7pt
\ligne{\hrulefill}
\vskip-3pt
\ligne{\zz \theblank\ #3\ \theblank} 
}} 

\newdimen\lalarge\lalarge=75pt
\vtop{ 
\begin{tab}\label{collis_minsky_r}
\leurre
Table of the collisions induced by the table of Minsky's Turing machine,
see Table~{\rm\ref{minskyUTM}}, impulsions to right.
\end{tab}
\vskip 5pt
\ligne{\hfill
\vtop{\leftskip 0pt\parindent 0pt\hsize=315pt
\ligne{\hfill
\hbox to\lalarge{\macrostepRR y 0 y y } \hbox to\lalarge{\macrostepRR y 1 A y }
\hbox to\lalarge{\macrostepRL y y y 0 } \hbox to\lalarge{\macrostepRR y A y 1 }
\hfill}
\vskip 9pt
\ligne{\hfill
\hbox to\lalarge{\macrostepRL 0 0 1 y } \hbox to\lalarge{\macrostepRR 0 1 A 0 }
\hbox to\lalarge{\macrostepRR 0 y y 0 } \hbox to\lalarge{\macrostepRR 0 A 1 0 }
\hfill}
\vskip 9pt
\ligne{\hfill
\hbox to\lalarge{\macrostepRL 1 0 1 A } \hbox to\lalarge{\macrostepRR 1 1 A 1 }
\hbox to\lalarge{\macrostepRR 1 y y 1 } \hbox to\lalarge{\macrostepRR 1 A 1 1 }
\hfill}
\vskip 9pt
\ligne{\hfill
\hbox to\lalarge{\macrostepRR A 0 y 1 } \hbox to\lalarge{\macrostepRR A 1 1 A }
\hbox to\lalarge{\macrostepRR A y 0 A } \hbox to\lalarge{\macrostepRR A A 0 y }
\hfill}
\vskip 9pt
}\hfill}
}

\vtop{ 
\begin{tab}\label{collis_minsky_l}
\leurre
Table of the collisions induced by the table of Minsky's Turing machine,
see Table~{\rm\ref{minskyUTM}}, impulsions to left.
\end{tab}
\vskip 5pt
\ligne{\hfill
\vtop{\leftskip 0pt\parindent 0pt\hsize=315pt
\ligne{\hfill
\hbox to\lalarge{\macrostepLL 0 y y 0 } \hbox to\lalarge{\macrostepLL 1 y 0 1 }
\hbox to\lalarge{\macrostepLL y y y 0 } \hbox to\lalarge{\macrostepLL A y y 1 }
\hfill}
\vskip 9pt
\ligne{\hfill
\hbox to\lalarge{\macrostepLR 0 0 y y } \hbox to\lalarge{\macrostepLR 1 0 A y }
\hbox to\lalarge{\macrostepLL y 0 y 0 } \hbox to\lalarge{\macrostepLR A 0 y 1 }
\hfill}
\vskip 9pt
\ligne{\hfill
\hbox to\lalarge{\macrostepLS 0 1 0 } \hbox to\lalarge{\macrostepLL 1 1 1 A }
\hbox to\lalarge{\macrostepLL y 1 1 y } \hbox to\lalarge{\macrostepLL A 1 A 1 }
\hfill}
\vskip 9pt
\ligne{\hfill
\hbox to\lalarge{\macrostepLR 0 A y 0 } \hbox to\lalarge{\macrostepLL 1 A B 1 }
\hbox to\lalarge{\macrostepLL y A A y } \hbox to\lalarge{\macrostepLL A A A 1 }
\hfill}
\vskip 9pt
\ligne{\hfill
\hbox to\lalarge{\macrostepLR 0 B y 1 } \hbox to\lalarge{\macrostepLR 1 B 1 A }
\hbox to\lalarge{\macrostepLR y B 0 A } \hbox to\lalarge{\macrostepLR A B 0 y }
\hfill}
\vskip 9pt
}
\hfill}
}

   Tables~\ref{collis_minsky_r} and~\ref{collis_minsky_l} represent the collisions of 
the various impulsions 
at the meeting with a symbol of Minsky's UTM in a cell as we have seen in
Sub-section~\ref{implement}. In the tables, note that there may be states with impulsions 
of both kinds. This is the case for states~2 and~7 exactly. Accordingly, there are two times 
four collisions devoted to each state, which means that we have a total of 36 collisions.

\newdimen\largebox\largebox=120pt
\setbox110=\vtop {\leftskip 0pt\parindent 0pt
\hsize=\largebox
\zz
\obeylines
\obeyspaces\global\let =\ \parskip=-2pt
 . |  . 0 1 y A B T 
--------------------
.. |  . . . . . . T
0. |  .   . . .   y 
1. |  . . . . .   1 
y. |  . . . .     y 
A. |  . . . .     1       
B. |  . . . . .           
T. |  T B 0 B A            
\par
}

\setbox112=\vtop {\leftskip 0pt\parindent 0pt
\hsize=\largebox
\zz
\obeylines
\obeyspaces\global\let =\ \parskip=-2pt
 0 |  . 0 1 y A B T 
--------------------
.0 |  0 y y 0 B 1 .   
00 |  1 A 0 1 y            
10 |  B   1 B   1 .        
y0 |  A   y   1   .        
A0 |  T   A y              
B0 |  y .   0     .        
T0 |  . .                  
\par
}

\setbox114=\vtop {\leftskip 0pt\parindent 0pt
\hsize=\largebox
\zz
\obeylines
\obeyspaces\global\let =\ \parskip=-2pt
 1 |  . 0 1 y A B T 
--------------------
.1 |  1 B 1 T 1 A .      
01 |  T   .       .        
11 |  0 0 y 0 B A .        
y1 |  y A 0 1 y            
A1 |  A A 0 A B y 1        
B1 |  1     A   y .        
T1 |  . B   . . .          
\par
}

\setbox116=\vtop {\leftskip 0pt\parindent 0pt
\hsize=\largebox
\zz
\obeylines
\obeyspaces\global\let =\ \parskip=-2pt
 y |  . 0 1 y A B T 
--------------------
.y |  y A T y 0 B .  
0y |  A 0 T 0 A 1 .        
1y |  y 1 A 1 1            
yy |  y 1 1 B 1 A .        
Ay |  A           .        
By |  0 y B   y            
Ty |  . B     . .          
\par
}

\setbox118=\vtop {\leftskip 0pt\parindent 0pt
\hsize=\largebox
\zz
\obeylines
\obeyspaces\global\let =\ \parskip=-2pt
 A |  . 0 1 y A B T 
--------------------
.A |  A 1 0 A 0 B .     
0A |  T 0   y T 0          
1A |  A 0     1   .        
yA |  0   B   1   .        
AA |  T 1   0 y            
BA |  1 A   A              
TA |  .     B   .          
\par
}

\setbox120=\vtop {\leftskip 0pt\parindent 0pt
\hsize=\largebox
\zz
\obeylines
\obeyspaces\global\let =\ \parskip=-2pt
 B |  . 0 1 y A B T 
--------------------
.B |  y y A T y y 1        
0B |  B y   A y            
1B |  T 0   A B .          
yB |  T 1   y .            
AB |  y A y A .            
BB |  A       1            
TB |  . .   0              
\par
}

\setbox122=\vtop {\leftskip 0pt\parindent 0pt
\hsize=\largebox
\zz
\obeylines
\obeyspaces\global\let =\ \parskip=-2pt
 T |  . 0 1 y A B T 
--------------------
.T |    0 1 y A B          
0T |  0     0              
1T |  1                    
yT |  y                    
AT |  A                    
BT |  B           .        
TT |  0                    
\par
}

\vtop{
\begin{tab}\label{tableLN}
Table of the cellular automaton with seven states as stated in~{\rm\cite{lindgren-nordahl}}.
\end{tab}
\ligne{\hfill\box110\hfill\box112\hfill\box114\hfill}
\vskip 10pt
\ligne{\hfill\box116\hfill\box118\hfill\box120\hfill}
\vskip 10pt
\ligne{\hfill\box122\hfill}
\vskip 10pt
}

\newdimen\collisbox\collisbox=350pt
\setbox110=\vtop {\leftskip 0pt\parindent 0pt
\hsize=\collisbox
\zz
\obeylines
\obeyspaces\global\let =\ \parskip=-2pt
     . 0 . T 0 .    . 1 . T 0 .    . . y . T 0 .    . A . T 0 . .    
     . 0 y 0 . .    . 1 1 0 . .    . . y y 0 . .    . A 1 0 . . .    
     . 0 0 A . .    . 1 0 B . .    . . y 1 A . .    . 0 A B . . .    
     . y y T . .    . B 1 B . .    . . T y A . .    . B 0 y . . .    
     . y . y T .    . A y T . .    . T y . 0 . .    . y 0 A . . .    
                    . A . y T .                     . A 1 T . . .    
                                                    . 0 1 1 T . .    
                                                    . y . . 1 T .    
\par
} 

\setbox112=\vtop {\leftskip 0pt\parindent 0pt
\hsize=\collisbox
\zz
\obeylines
\obeyspaces\global\let =\ \parskip=-2pt
     . 0 . T 1 .    . . 1 . T 1 .    . . y . T 1 .    . . . A . T 1 .    
     . 0 y 1 . .    . . 1 1 1 . .    . . y y 1 . .    . . . A 1 1 . .     
     . 0 T y . .    . . 1 y 0 . .    . . y 1 y . .    . . . 0 0 0 . .     
     . . 0 . . .    . . T 1 A . .    . . T 1 y . .    . . . y A 1 . .     
                    . T 1 . A . .    . T 1 . y . .    . . . 0 B A . .     
                                                      . . . 1 y 1 . .     
                                                      . . . T A y . .     
                                                      . . T A B A . .     
                                                      . T A . . 1 . .     
\par}

\setbox114=\vtop {\leftskip 0pt\parindent 0pt
\hsize=\collisbox
\zz
\obeylines
\obeyspaces\global\let =\ \parskip=-2pt
     . . 0 . T y .    . . 1 . T y .    . . y . T y .    . . . A . T y .    
     . . 0 y y . .    . . 1 1 y . .    . . y y y . .    . . . A 1 y . .    
     . . 0 0 y . .    . . 1 0 y . .    . . y B y . .    . . . 0 A y . .    
     . . y 1 A . .    . . B B A . .    . . B y 0 . .    . . . B y A . .    
     . . T y A . .    . . y 1 1 . .    . . T y A . .    . . . T y 0 . .    
     . T y . 0 . .    . . T 0 0 . .    . T y . 0 . .    . . T y B A . .    
                      . T 0 . 1 . .                     . T y . . 1 . .    
\par
}                                                                                                 
                                                                                                 
\setbox116=\vtop {\leftskip 0pt\parindent 0pt
\hsize=\collisbox
\zz
\obeylines
\obeyspaces\global\let =\ \parskip=-2pt
     . 0 . T A .    . . . 1 . T A .    . . . y . T A .    . . . A . T A .    
     . 0 y A . .    . . . 1 1 A . .    . . . y y A . .    . . . A 1 A . .    
     . 0 A 0 . .    . . . 1 B A . .    . . . y 1 0 . .    . . . 0 B A . .    
     . B 0 T . .    . . . A B 1 . .    . . . T A B . .    . . . 1 y 1 . .    
     . y . 0 T .    . . . B y 1 . .    . . T A . y . .    . . . T A y . .    
         . . 0 T    . . . T B y . .                  .    . . T A B A . .    
     . . . y . .    . . T B 0 0 . .    T A . . . y . .    . T A . . 1 . .    
     . . . y . .    . T B . . 1 . .    A . . . . y . .    
\par
}

\setbox118=\vtop {\leftskip 0pt\parindent 0pt
\hsize=\collisbox
\zz
\obeylines
\obeyspaces\global\let =\ \parskip=-2pt
      . 0 . T B .    . 1 . T B .    . y . T B .    . A . T B . 
      . 0 y B . .    . 1 1 B . .    . y y B . .    . A 1 B . . 
      . 0 1 T . .    . 1 A T . .    . y A T . .    . 0 y T . . 
      . y . 1 T .    . 1 . A T .    . 0 . A T .    . 0 . y T . 
\par
}

\setbox120=\vtop {\leftskip 0pt\parindent 0pt
\hsize=\collisbox
\zz
\obeylines
\obeyspaces\global\let =\ \parskip=-2pt
     . 0 T . 0 .    . 0 T . 1 . .    . 0 T . y . . .    . 0 T . A . .    
     . . 0 B 0 .    . . 0 0 1 . .    . . 0 B y . . .    . . 0 A A . .    
     . . 1 y y .    . . y 0 T . .    . . 1 A 0 . . .    . . B T T . .    
     . . T 1 y .    . . A . 0 T .    . . 1 0 T . . .    . . 1 . 0 T .    
     . T 1 . y .                     . . B . 0 T . . 
                                     . . y . . 0 T . 
\par
}

\setbox122=\vtop {\leftskip 0pt\parindent 0pt
\hsize=\collisbox
\zz
\obeylines
\obeyspaces\global\let =\ \parskip=-2pt
     . . 1 T . 0 .    . 1 T . 1 . .    . 1 T . y . .    . 1 T . A . .    
     . . . 1 B 0 .    . . 1 0 1 . .    . . 1 B y . .    . . 1 A A . .    
     . . . A 0 y .    . . B 1 T . .    . . A A 0 . .    . . 1 1 T . .    
     . . . 1 y A .    . . A . 1 T .    . . 0 1 T . .    . . 1 . 1 T .   
     . . . T 1 0 .                     . . y . 1 T .    
     . . T 1 B B .    
     . T 1 . . A .    
\par
}                                                                                                 

\setbox124=\vtop {\leftskip 0pt\parindent 0pt
\hsize=\collisbox
\zz
\obeylines
\obeyspaces\global\let =\ \parskip=-2pt
     . y T . 0 . .    . y T . 1 . .    . y T . y .    . y T . A . . .    
     . . y B 0 . .    . . y 0 1 . .    . . y B y .    . . y A A . . .    
     . . B 1 y . .    . . A y T . .    . . B y 0 .    . . 0 1 T . . .    
     . . A A y . .    . . A . y T .    . . T y A .    . . y . 1 T . .    
     . . 0 0 A . .                     . T y . 0 .    
     . . y y T . .    
     . . y . y T .    
\par
}

\setbox126=\vtop {\leftskip 0pt\parindent 0pt
\hsize=\collisbox
\zz
\obeylines
\obeyspaces\global\let =\ \parskip=-2pt
     . A T . 0 . .    . A T . 1 . .    . A T . y . .    . A T . A . .    
     . . A B 0 . .    . . A 0 1 . .    . . A B y . .    . . A A A . .    
     . . B A y . .    . . 1 A T . .    . . B A 0 . .    . . 0 y T . .    
     . . y A A . .    . . 1 . A T .    . . y A T . .    . . 0 . y T .    
     . . 0 1 T . .    . . 1 . . A T    . . 0 . A T . 
     . . y . 1 T .    . . 1 . . . A    
\par
}

\vtop{
\begin{tab}\label{execright}
\leurre
Execution of the rules for the collisions of a right-hand side impulsion with
a Turing letter. Note that the blank is represented by~{\tt .}
\end{tab}
\ligne{\hfill\box120\hfill}
\vskip 5pt
\ligne{\hfill\box122\hfill}
\vskip 5pt
\ligne{\hfill\box124\hfill}
\vskip 5pt
\ligne{\hfill\box126\hfill}
\vskip 5pt
}

   Table~\ref{tableLN} gives the rules of~$\cal LN$ as described in~\cite{lindgren-nordahl}.
Our presentation is slightly different. The table in~\ref{tableLN} is split into
seven small tables labelled by a state of~$\cal LN$. The sub-table~{\tt x} 
gives all rules of the form \hbox{\tt l x r $\rightarrow$ y}, where {\tt x} is the state
of the cell, {\tt l$,$ r} the state of its left-, right-hand side neighbour respectively,
and {\tt y} is the new state of the cell after the application of the rule. 
The entry~{\tt y} is at the intersection of the row \hbox{\tt lx} and the column~{\tt r}.

\vtop{
\begin{tab}\label{execleft}
\leurre
Execution of the rules for the collisions of a left-hand side impulsion with
a Turing letter. Here too, the blank is represented by~{\tt .}
\end{tab}
\ligne{\hfill\box110\hfill}
\vskip 5pt
\ligne{\hfill\box112\hfill}
\vskip 5pt
\ligne{\hfill\box114\hfill}
\vskip 5pt
\ligne{\hfill\box116\hfill}
\vskip 5pt
\ligne{\hfill\box118\hfill}
\vskip 5pt
}

   Tables~\ref{execright} and~\ref{execleft} reproduce those of~\cite{lindgren-nordahl}.
They were produced by a computer program mimicking~$\cal LN$, applying the rules 
of Table~\ref{tableLN}. This allows us to check that the collisions depicted by
Tables~\ref{collis_minsky_r} and~\ref{collis_minsky_l} can be performed according to 
the guidelines of
Subsection~\ref{guideline}.

\def\LN{\hbox{$\cal LN$}}

\section{Strongly universal cellular automata on a one-dimensional line}
\label{strong}

    In this section, we complete the description of~\LN{} so that it now satisfies the
requirements of strong universality as defined in Section~\ref{intro}.

   As mentioned in Section~\ref{exec}, \LN{} is not yet strongly universal: it works
in an infinite non-blank background. The idea is to start with a finite initial
configuration, but we modify it in such a way that it constructs the background
so that each time the Turing head goes out from the current configuration to the right,
it meets the required pattern of the background.

   We can improve~\LN{} by providing a construction of the background
during the computation of the automaton. There is a very simple way to do that
by using three additional states, say {\tt \#}, {\tt \$} and {\tt \&}. 

   From the explanations of Sub-subsection~\ref{minskystrong}, as Minsky's UTM
simulates a tag-system, there is a square~$c$ of the Turing tape such that the head of the
machine never goes to the left of~$c$. And so, we can consider that all cells to the
left of~$c$ are blank. As the initial configuration of finite, there is a cell~$d$
such that all cells on the right-hand side of~$d$ are blank. And so we can choose~$d$
so that the interval $[c,d]$ contains the encoding of the initial configuration of
Minsky's UTM.
From Sub-subsection~\ref{minskystrong} again, we remark that when 
Minsky's UTM halts, the head is in between the final tagged word and the productions,
looking at a~{\tt 0}.

    Consider the leftmost~{\tt 0} which is on the right-hand side of~$d$. 
Replace~{\tt 0} by~{\tt \#}. Table~\ref{back1} provides us with a scheme of 
execution which shows how it is possible to construct the background needed for \LN{}
during its computation :

\newdimen\largeend\largeend=130pt
\setbox110=\vtop{\leftskip 0pt\parindent 0pt\hsize=\largesitu
\zz
\obeylines
\obeyspaces\global\let =\ \parskip=-2pt
 . \# . . . . . . .
 . 0 \$ . . . . . .
 . 0 . \& . . . . .
 . 0 . . \# . . . .
\par}
\vtop{
\begin{tab}\label{back1}
\leurre
Using three states to construct the background for \LN.
\end{tab}
\ligne{\hfill\box110\hfill}
\vskip 7pt
}

   However, this scheme is not suitable for our purpose: when the Turing machine halts,
a signal should be sent in order to reach {\tt \$$,$ \&} or~{\tt \#} in order to stop the
construction of the background for~\LN. But this cannot be achieved: The maximal speed of a
signal is~1 and that of the above symbols in the scheme of Table~\ref{back1} is also~1, so
that it can never be reached. We use another pattern given by Table~\ref{back2} which 
makes use of two additional states only.
   
\setbox110=\vtop{\leftskip 0pt\parindent 0pt\hsize=\largesitu
\zz
\obeylines
\obeyspaces\global\let =\ \parskip=-2pt
. . U U . . . . . .
. . U U B . . . . .
. . U U U . . . . .
. . U U U B . . . .
. . U U U U . . . .
. . U U U U B . . .
. . U U U U U . . .
\par}
\vtop{
\begin{tab}\label{back2}
\leurre
Using an appropriate pattern to construct the background needed by\LN.
\end{tab}
\ligne{\hfill\box110\hfill}
}
\vskip 7pt
   The advantage of this new pattern is that its speed is~$\displaystyle{1\over2}$,
so that it will be reached by a signal sent later at speed~1. But the pattern has also
to be not too slow: it must provide the background in advance and so, it must
go faster than the advance of the computation itself. We can see in Tables~\ref{execleft}
and~\ref{execright} that the speed of the computation of~\LN{} oscillates
between $\displaystyle{1\over7}$ and $\displaystyle{1\over3}$.

   This pattern leaves a whole interval of~{\tt U}'s growing to the right-hand side
and which starts from the right-hand side neighbour of the cell~$d$.

   We also have to implement the signal which will reach the pattern constructing the
background. An additional state~{\tt 3} is used to erase the {\tt U}'s until reaching 
the last occurrence of~{\tt B} which is erased too. We call it the {\bf stopping signal}. 
Note that {\tt 3} is raised by the neighbouring {\tt 0Ty} of~\LN. We simply replace the
rule \hbox{\tt OTy $\rightarrow$ 0} by the rule \hbox{\tt OTy $\rightarrow$ 3}. But there is
an additional constraint: signal~{\tt 3} starts from the left-hand end of the final tagged
word and it has to reach the right-hand side end of the configuration. Accordingly,
it must cross the tagged word without damaging it. For this purpose, we introduce two more 
states: {\tt 4} and~{\tt V}. States~{\tt 3} and~{\tt 4} replace and then restore {\tt 0}, 
{\tt 1} respectively, while {\tt U} and~{\tt V} perform the same operations on~{\tt y} 
and~{\tt A} respectively. 

   Now, when the Turing head starts to examine the cell~$d$, the rules we have seen
in Sub-subsection~\ref{trueexec} do not apply. So that we have to append new rules 
for the impulsions going to the right and meeting a~{\tt 0}: now, {\tt U} has the 
meaning of a~{\tt 0}. Another state is needed, {\tt 4}, also interpreted as~{\tt 0}.
The role of state~{\tt 4} is to create the two blanks which separate the Turing symbols 
on the tape. Table~\ref{new_collis} shows how this works. Table~\ref{table_reg} also 
contains the corresponding rules and Table~\ref{new_collis} was obtained by a computer
applying the rules of Table~\ref{table_reg}.

\advance \largeend by 2.3pt
\setbox110=\vtop{\leftskip 0pt\parindent 0pt\hsize=\largeend
\zz
\obeylines
\obeyspaces\global\let =\ \parskip=-2pt
 . |  . 0 1 y A B T 3 4 U V
---------------------------
.. |  . . . . . . T . . . .
0. |  .   . . .   y . .   .
1. |  . . . . .   1 .      
y. |  . . . .     y .      
A. |  . . . .     1 .      
B. |  . . . . .     .      
T. |  T B 0 B A         B  
3. |  3 3 3 3 3            
4. |  3                    
U. |  B                    
V. |  3                    
\par} 
\setbox112=\vtop{\leftskip 0pt\parindent 0pt\hsize=\largeend
\zz
\obeylines
\obeyspaces\global\let =\ \parskip=-2pt
 0 |  . 0 1 y A B T 3 4 U V
---------------------------
.0 |  0 y y 0 B 1 . 0      
00 |  1 A 0 1 y            
10 |  B   1 B   1 .        
y0 |  A   y   1   .        
A0 |  T   A y              
B0 |  y .   0     .        
T0 |  . .                  
30 |  3                    
40 |                       
U0 |                       
V0 |                       
\par} 
\setbox114=\vtop{\leftskip 0pt\parindent 0pt\hsize=\largeend
\zz
\obeylines
\obeyspaces\global\let =\ \parskip=-2pt
 1 |  . 0 1 y A B T 3 4 U V
---------------------------
.1 |  1 B 1 T 1 A . 1      
01 |  T   .       .        
11 |  0 0 y 0 B A .        
y1 |  y A 0 1 y            
A1 |  A A 0 A B y 1        
B1 |  1     A   y .        
T1 |  . B   . . .          
31 |  4                    
41 |                       
U1 |                       
V1 |                       
\par} 
\setbox116=\vtop{\leftskip 0pt\parindent 0pt\hsize=\largeend
\zz
\obeylines
\obeyspaces\global\let =\ \parskip=-2pt
 y |  . 0 1 y A B T 3 4 U V
---------------------------
.y |  y A T y 0 B . y      
0y |  A 0 T 0 A 1 .     A  
1y |  y 1 A 1 1       y y  
yy |  y 1 1 B 1 A .     y  
Ay |  A           .   A A  
By |  0 y B   y            
Ty |  . B     . .          
3y |  U                    
4y |                       
Uy |                       
Vy |                       
\par} 
\setbox118=\vtop{\leftskip 0pt\parindent 0pt\hsize=\largeend
\zz
\obeylines
\obeyspaces\global\let =\ \parskip=-2pt
 A |  . 0 1 y A B T 3 4 U V
---------------------------
.A |  A 1 0 A 0 B . A      
0A |  T 0   y T 0          
1A |  A 0     1   .        
yA |  0   B   1   .   0    
AA |  T 1   0 y       T    
BA |  1 A   A              
TA |  .     B   .          
3A |  V                    
4A |                       
UA |                       
VA |                       
\par} 
\setbox120=\vtop{\leftskip 0pt\parindent 0pt\hsize=\largeend
\zz
\obeylines
\obeyspaces\global\let =\ \parskip=-2pt
 B |  . 0 1 y A B T 3 4 U V
---------------------------
.B |  y y A T y y 1        
0B |  B y   A y         y  
1B |  T 0   A B .       0  
yB |  T 1   y .         1  
AB |  y A y A .         A  
BB |  A       1            
TB |  . .   0              
3B |                       
4B |  .                    
UB |  U                    
VB |  3                    
\par} 
\setbox122=\vtop{\leftskip 0pt\parindent 0pt\hsize=\largeend
\zz
\obeylines
\obeyspaces\global\let =\ \parskip=-2pt
 T |  . 0 1 y A B T 3 4 U V
---------------------------
.T |    0 1 y A B          
0T |  0     3              
1T |  1                    
yT |  y                    
AT |  A                    
BT |  B           .        
TT |  0                    
3T |                       
4T |                       
UT |                       
VT |                       
\par} 
\setbox124=\vtop{\leftskip 0pt\parindent 0pt\hsize=\largeend
\zz
\obeylines
\obeyspaces\global\let =\ \parskip=-2pt
 3 |  . 0 1 y A B T 3 4 U V
---------------------------
.3 |  0 . . . .         .  
03 |  .                    
13 |  .                    
y3 |  .                    
A3 |  .                    
B3 |  .                    
T3 |                       
33 |                       
43 |                       
U3 |                       
V3 |                       
\par} 
\setbox126=\vtop{\leftskip 0pt\parindent 0pt\hsize=\largeend
\zz
\obeylines
\obeyspaces\global\let =\ \parskip=-2pt
 4 |  . 0 1 y A B T 3 4 U V
---------------------------
.4 |  1         .       .  
04 |                       
14 |                       
y4 |                    .  
A4 |                    .  
B4 |                       
T4 |                       
34 |                       
44 |                       
U4 |                       
V4 |                       
\par} 
\setbox128=\vtop{\leftskip 0pt\parindent 0pt\hsize=\largeend
\zz
\obeylines
\obeyspaces\global\let =\ \parskip=-2pt
 U |  . 0 1 y A B T 3 4 U V
---------------------------
.U |  V                 U  
0U |                       
1U |                       
yU |                    4  
AU |                       
BU |                    y  
TU |                       
3U |            4       3  
4U |  4                 .  
UU |  U         U       U  
VU |                       
\par} 
\setbox130=\vtop{\leftskip 0pt\parindent 0pt\hsize=\largeend
\zz
\obeylines
\obeyspaces\global\let =\ \parskip=-2pt
 V |  . 0 1 y A B T 3 4 U V
---------------------------
.V |  A         y          
0V |                       
1V |                       
yV |                       
AV |                       
BV |                       
TV |                       
3V |                       
4V |                       
UV |                       
VV |                       
\par}
\vtop{
\begin{tab}\label{table_reg}
\leurre
Table of the rules to complete the cellular automaton \LN{} into the rules
for \LN$_{11}$.
\end{tab}
\ligne{\hfill\box110\hfill\box112\hfill\box114\hfill}
\vskip 10pt
\ligne{\hfill\box116\hfill\box118\hfill\box120\hfill}
\vskip 10pt
\ligne{\hfill\box122\hfill\box124\hfill\box126\hfill}
\vskip 10pt
\ligne{\hfill\box128\hfill\box130\hfill}
\vskip 10pt
}

   Table~\ref{crosstag} shows how the states are used for this crossing. For a technical reason,
the crossing of a~{\tt y} is slightly more complex: as we need the rule
\hbox{\tt U . . $\rightarrow$ B}, we cannot have a rule \hbox{\tt . U . $\rightarrow$ y}
as the other patterns of the Table would require. Instead, we use the simultaneous
occurrence of~{\tt V} and~{\tt B} to avoid any conflict with the rules defining the
motion indicated by Table~\ref{back2}.

   The rules given in Table~\ref{table_reg} perform the computations given in
Tables~\ref{back2} and~\ref{crosstag}. They also allow to perform the computations
given in Table~\ref{stopping} which show how the stopping signal halts the progression
of the background. 

\setbox110=\vtop{\leftskip 0pt\parindent 0pt\hsize=180pt
\zz
\obeylines
\obeyspaces\global\let =\ \parskip=-2pt
3 . y . . y . . A . . 0 . . 1 . . 
. 3 y . . y . . A . . 0 . . 1 . . 
. . U . . y . . A . . 0 . . 1 . .
. . V B . y . . A . . 0 . . 1 . .
. . y 3 . y . . A . . 0 . . 1 . .
. . y . 3 y . . A . . 0 . . 1 . .
. . y . . U . . A . . 0 . . 1 . .
. . y . . V B . A . . 0 . . 1 . .
. . y . . y 3 . A . . 0 . . 1 . .
. . y . . y . 3 A . . 0 . . 1 . .
\par}
\setbox112=\vtop{\leftskip 0pt\parindent 0pt\hsize=180pt
\zz
\obeylines
\obeyspaces\global\let =\ \parskip=-2pt
. . y . . y . 3 A . . 0 . . 1 . .
. . y . . y . . V . . 0 . . 1 . .
. . y . . y . . A 3 . 0 . . 1 . .
. . y . . y . . A . 3 0 . . 1 . . 
. . y . . y . . A . . 3 . . 1 . .
. . y . . y . . A . . 0 3 . 1 . .
. . y . . y . . A . . 0 . 3 1 . .
. . y . . y . . A . . 0 . . 4 . .
. . y . . y . . A . . 0 . . 1 3 .
. . y . . y . . A . . 0 . . 1 . 3
\par}
\vtop{
\begin{tab}\label{crosstag}
\leurre
The progression of the stopping signal across the final tagged word.
The first line of the right-hand side table reproduce the last line of the left-hand side
one for clarity reason.
\end{tab}
\ligne{\hfill\box110\hfill\box112\hfill}
\vskip 5pt
}

Let us now call~\LN$_{11}$ the cellular automaton with 11~states whose rules
are given in Table~\ref{table_reg}. As Table~\ref{tableLN}, the table is organized in 
two-dimensional sub-tables, according to the same pattern.

   It can be noticed that there are a lot of empty entries. This corresponds to the fact that
the table of~\LN$_{11}$ is constructed for specific configurations attached to our purpose.
The configurations we consider are assumed to encode the current configuration of a Turing
machine so that not any triple of the states occurs during these configurations. This 
particularity will be useful for the adaptation of~\LN$_{11}$ to a strongly universal
cellular automaton in the hyperbolic tilings we shall consider in Section~\ref{extent}.

\setbox110=\vtop{\leftskip 0pt\parindent 0pt\hsize=100pt
\zz
\obeylines
\obeyspaces\global\let =\ \parskip=-2pt
. 3 U U U U . . . . .
. . 3 U U U B . . . .
. . . 3 U U U . . . .
. . . . 3 U U B . . .
. . . . . 3 U U . . .
. . . . . . 3 U B . .
. . . . . . . 4 U . .
. . . . . . . . 4 B .
. . . . . . . . . . .
\par}
\vtop{
\begin{tab}\label{stopping}
When the stopping signal arrives to the end of the zone of~{\tt U}'s.
\end{tab}
\ligne{\hfill\box110\hfill}
\vskip 10pt
}

\setbox110=\vtop{\leftskip 0pt\parindent 0pt\hsize=100pt
\zz
\obeylines
\obeyspaces\global\let =\ \parskip=-2pt
. y T . U U U . . . .
. . y B U U U B . . .
. . B 1 y U U U . . .
. . A A y 4 U U B . .
. . 0 0 A . . U U . .
. . y y T . . U U B .
. . y . y T . U U U .
. . y . . y B U U U B
\par}
\setbox112=\vtop{\leftskip 0pt\parindent 0pt\hsize=100pt
\zz
\obeylines
\obeyspaces\global\let =\ \parskip=-2pt
. 0 T . U U U . . . 
. . 0 B U U U B . .
. . 1 y y U U U . .
. . T 1 y 4 U U B .
. T 1 . y . . U U .
\par}
\setbox114=\vtop{\leftskip 0pt\parindent 0pt\hsize=100pt
\zz
\obeylines
\obeyspaces\global\let =\ \parskip=-2pt
. . 1 T . U U U . . .
. . . 1 B U U U B . .
. . . A 0 y U U U . .
. . . 1 y A 4 U U B .
. . . T 1 0 . . U U .
. . T 1 B B . . U U B
. T 1 . . A . . U U U
\par}
\setbox116=\vtop{\leftskip 0pt\parindent 0pt\hsize=100pt
\zz
\obeylines
\obeyspaces\global\let =\ \parskip=-2pt
. A T . U U U . . . .
. . A B U U U B . . .
. . B A y U U U . . .
. . y A A 4 U U B . .
. . 0 1 T . . U U . .
. . y . 1 T . U U B .
. . y . . 1 B U U U .
\par}
\vtop{
\begin{tab}\label{new_collis}
Execution of the rules of Table~{\rm\ref{table_reg}} for collisions
with the zone of~{\tt U}'s.
\end{tab}
\ligne{\hfill\box110\hfill\box112\hfill\box114\hfill\box116\hfill}
\vskip 10pt
}

Accordingly, \LN$_{11}$ allows us to prove:

\begin{thm}\label{strong1D}
There is a deterministic cellular automaton on the line, with the standard neighbourhood
and $11$~states which is strongly universal.
\end{thm}

We have an interesting corollary:

\begin{cor}\label{haltundec}
There is a deterministic cellular automaton on the line, with the standard neighbourhood
and $9$~states whose halting problem when starting from a finite configuration
is undecidable.
\end{cor}

Proof. If we look only at the halting problem, when the simulated machine halts, it is no more
necessary to keep the final result of the tag-system. This allows us to spare the additional
states~{\tt 3} and~{\tt V}. However, we need a stopping signal. This can be performed by
state~{\tt 4}. We replace the rule \hbox{\tt OTy $\rightarrow$ 3} 
by \hbox{\tt OTy $\rightarrow$ 4}. Later, we can decide that state {\tt 4} destroys any symbol.
The rules involving~{\tt 3} and~{\tt V} are cancelled. We have the following new rules:
\vskip 5pt
\ligne{\hfill\tt \theblank{}4$\alpha\rightarrow$ \theblank\hfill
4$\alpha$\theblank{} $\rightarrow$ 4\hfill
\theblank\theblank4 $\rightarrow$ \theblank\hfill
4\theblank\theblank{} $\rightarrow$ \theblank\hfill
4\theblank U $\rightarrow$ 4\hfill
\theblank 4B $\rightarrow$\theblank\hfill
4B \theblank{} $\rightarrow$ \theblank,\hfill}
\vskip 5pt
\noindent
where $\alpha$~stands for {\tt 0$,$ 1$,$ y$,$} or~{\tt A}, as {\tt B} and~{\tt T} disappeared
and
as far as the simulated machine halted. We keep the rules 
\hbox{\tt\theblank 4U $\rightarrow$ \theblank} and
\hbox{\tt 4UU $\rightarrow$ \theblank}.
\qed
  

\section{Strongly universal cellular automata in hyperbolic spaces}
\label{extent}

   We shall start from the automaton \LN{} to which we appended the states~{\tt U}
and{\tt 4} with the rules producing Tables~\ref{back2} and~\ref{new_collis}. This
automaton, call it \LN$_9$ has 9~states but it does not halt when the simulated machine
halts.

   The idea is to implement \LN$_9$ on a line which is continued in such a way that,
taking advantage of the hyperbolic structure, we can simplify the process which
halts the computation. We shall take the approach given in~\cite{mmstrong}.
There, the construction of a line supporting the computation of $1D$-cellular automaton 
is performed starting from a finite initial segment of the cells which contains the 
implementation of the finite initial configuration of the Turing machine. Such a construction 
can be viewed as an independent cellular automaton~$\cal P$. The simple juxtaposition of both 
automata, $\cal P$ and the implementation of~\LN$_9$, leads to a new cellular automaton. 
With some minor tuning, this will
allow us to prove the results claimed in the introduction, see Theorem~\ref{stronghyp10}.

In Sub-section~\ref{hyptilings}, a minimal presentation of the tilings in which we implement 
cellular automata is given. In Sub-section~\ref{continuation} we remind the reader the main 
lines of the construction of~$\cal P$. In Sub-section~\ref{stronghyp}, the combination 
of~$\cal P$ with~\LN$_9$ is performed, proving 
the results stated
by Theorem~\ref{stronghyp11} and~\ref{stronghyp10}. 

\subsection{Tilings of the hyperbolic plane and cellular automata}
\label{hyptilings}

   It is not possible to remind here all the properties from hyperbolic geometry needed
for the paper. I very sketchily remind here what is needed to understand the
arguments developed in this section. The reader is referred to~\cite{mmbook1,mmbook3} for 
a more informative introduction to the field, especially suited for the paper. 

   In this paper, we again use Poincar\'e's disc model of the hyperbolic plane as well as
its generalization to the $3D$~ball as a model of the hyperbolic $3D$-space.

   Poincar\'e's disc model is illustrated by Figure~\ref{poincaredisc}. The points of the
hyperbolic plane are the points of a fixed open disc~$D$ whose border~$\partial D$ is drawn in 
the figure. The points of~$\partial D$ are called the {\bf points at infinity} and they do
not belong to the hyperbolic plane. The lines are represented by the traces in~$D$ of diameters
or of circles which are orthogonal to~$\partial D$. In the figure, a line~$\ell$ is represented
together with a point~$A$ out of~$\ell$. Three kinds of lines are represented in the figure.
The first kind of lines is represented by~$s$: they all cut~$\ell$ and are thus called
{\bf secant} with~$\ell$. Two lines constitute the second kind: they share with~$\ell$
a point at infinity. They are illustrated by~$p$ and~$q$ in the figure. The lines~$p$ and~$q$
pass through~$A$ and meet~$\ell$ exactly at the points at infinity~$P$ and~$Q$ respectively.
Accordingly, they do not meet~$\ell$ in the hyperbolic plane and are therefore called
{\bf parallel} to~$\ell$. And so, in the model, through each point out of a line, we can
draw two distinct parallels to the line. Now we have a third kind of lines which has no 
counterpart in Euclidean geometry. It consists of the lines passing through~$A$ which do not 
meet~$\ell$ at all: neither in the hyperbolic plane, nor at infinity, nor outside the closure 
of~$D$ in the model. These lines are illustrated in the figure by~$m$ and are called 
{\bf non-secant} with~$\ell$. Non-secant lines are characterized by the fact that they have
a unique common perpendicular.

\vtop{
\vskip 8pt
\ligne{\hfill\includegraphics[scale=1]{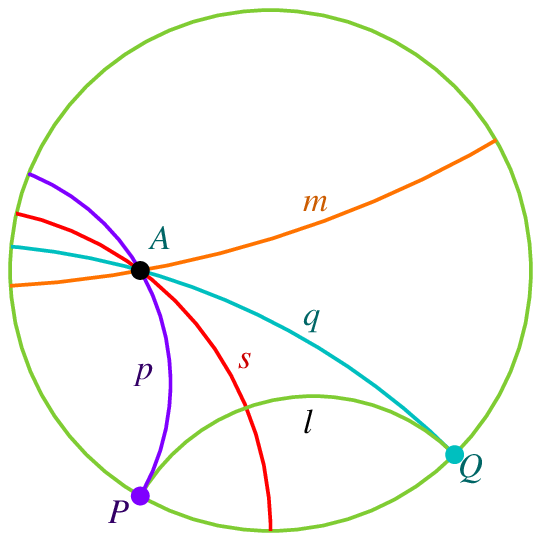}\hfill}
\begin{fig}\label{poincaredisc}
\vspace{-8pt}
\leurre
Poincar\'e's disc. The three kinds of lines passing through a point out of a given line.
\end{fig}
\vskip 8pt
}

A tessellation is defined by the following process. We start from a convex regular polygon~$P$,
the {\bf basis} and we replicate~$P$ by reflection in its sides and, recursively, the 
images in their sides. We call~$P$ and these images of~$P$ {\bf copies} of~$P$. We
say that this process defines a {\bf tessellation} if and only if the copies do not
overlap and they cover all points of the plane. In the Euclidean plane, up to similarity,
there are three tessellations exactly. The bases are the square, the regular hexagon and
the equilateral triangle.

   The important news is that in the hyperbolic plane, there are infinitely many 
tessellations. A basis~$P$ of such a tessellation is defined by two numbers: the number~$p$
of sides of~$P$ and the number~$q$ of copies of~$P$ which can be put around a point~$A$ in order
to cover a neighbourhood of~$A$ with no overlapping. The tessellation itself is
denoted by~$\{p,q\}$. A fundamental theorem proved by Poincar\'e says that $P$
generates a tessellation of he hyperbolic plane if and only if 

\ligne{\hfill$\displaystyle{{1\over p}+{1\over q} < {1\over2}}$.\hfill($\star$)\hskip 20pt}

   Note that we recover the Euclidean tessellations by replacing~$<$ by~$=$ in~$(\star)$.

In the hyperbolic plane, the smallest values for~$q$ are~3 and~4 and the smallest 
corresponding values of~$p$ such that $(\star)$ holds are~7 and~5 respectively. The 
corresponding tilings are called {\bf heptagrid} and {\bf pentagrid} respectively,
see Figure~\ref{hypgrids}.

    As mentioned in the introduction of the sub-section, Poincar\'e's disc model generalizes
to the hyperbolic $3D$ space and, in fact, to any higher dimension. Planes are the trace
in the unit open ball~$B$ of diametral planes or spheres which are orthogonal to the
border~$\partial B$ of~$B$. Lines are intersections of secant planes. There are four
tessellations in the hyperbolic $3D$ space and we shall consider one of them, only,
the {\bf dodecagrid}. It is based on what is called Poincar\'e's dodecahedron which is 
constructed upon the convex rectangular regular pentagon in the same way as a Euclidean 
dodecahedron is constructed upon a convex regular pentagon. It turns out that this
dodecahedron tiles the hyperbolic $3D$ space.

\vtop{
\vskip 8pt
\ligne{\hfill\includegraphics[scale=0.8]{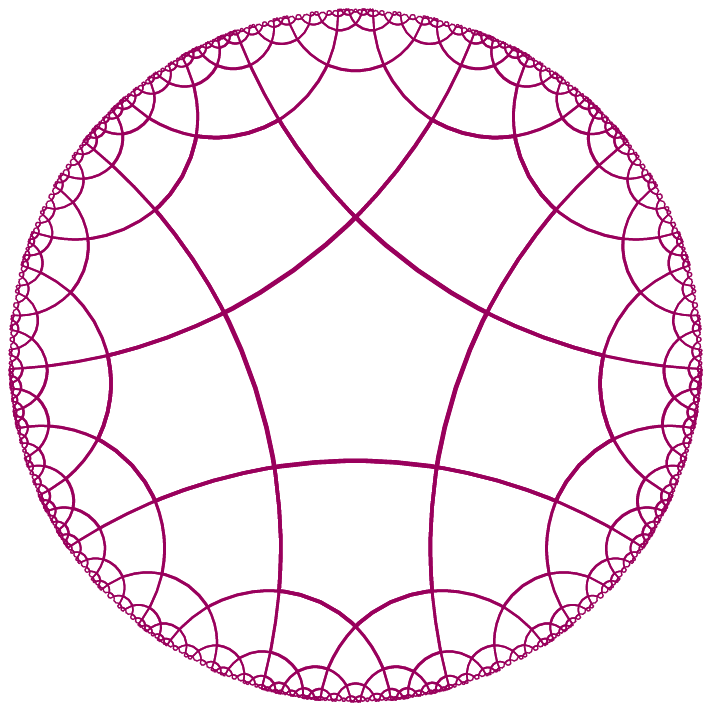}\hfill
\includegraphics[scale=0.8]{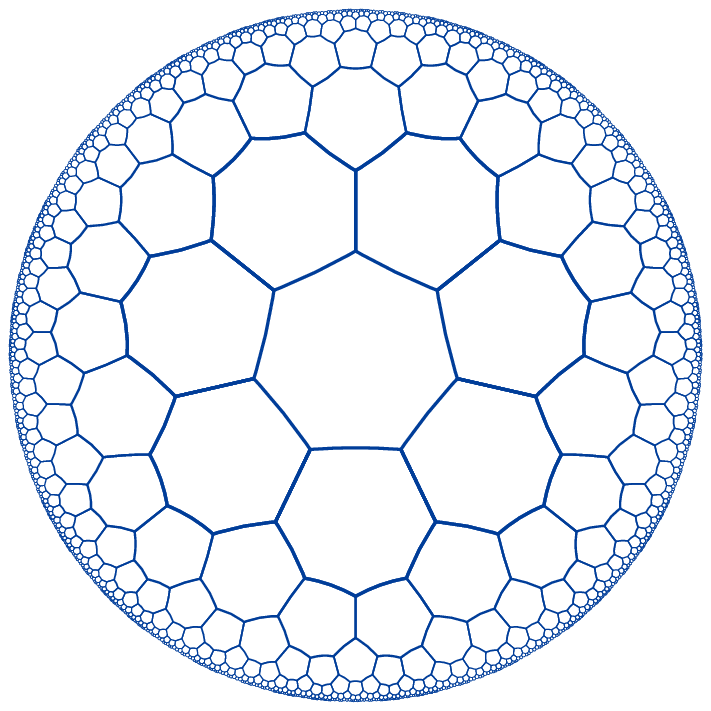}\hfill}
\begin{fig}\label{hypgrids}
\leurre
Hyperbolic tessellations. To left: the pentagrid; to right: the heptagrid.
\end{fig}
\vskip 8pt
}

    Figure~\ref{dodecviews} illustrates how we represent dodecahedra and the dodecagrid
in this setting. For our purpose, we need to fix two perpendicular planes~$H$ and~$V$
which support faces of dodecahedra of the tiling. 

\vtop{
\ligne{\hfill\includegraphics[scale=0.8]{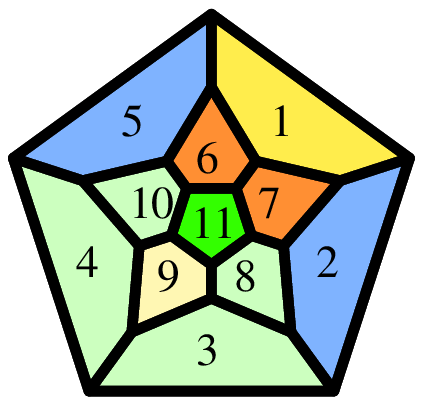}\hfill
\includegraphics[scale=1]{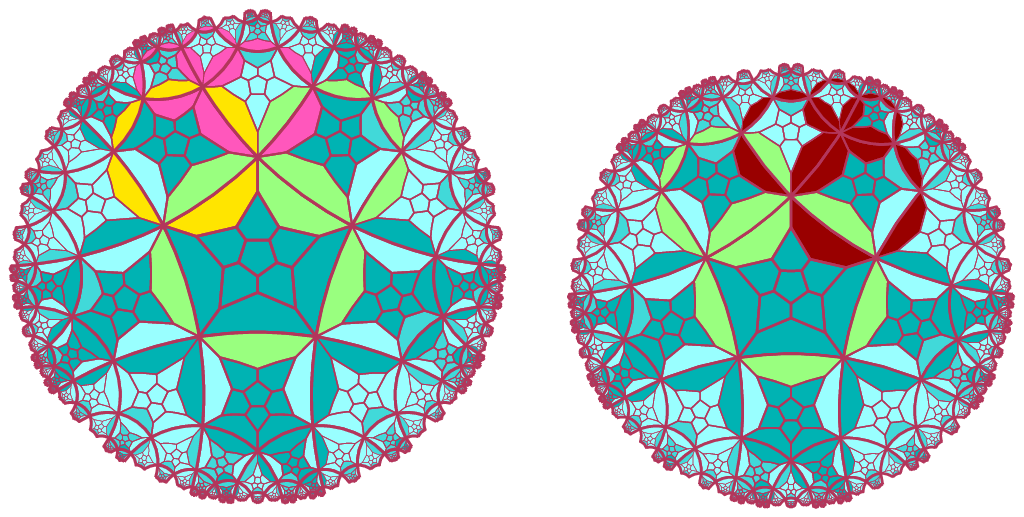}\hfill}
\vspace{-10pt}
\begin{fig}\label{dodecviews}
\leurre
To left:
the numbering of the faces of a cell. Face~$0$ is on the plane~$\cal H$, face~$1$ is on
the plane~$\cal V$. By construction, the non-blank faces are~$0$, $1$, $6$, $7$. In a cell of
the yellow line, faces~$2$ and~$5$ may be blank but no both at the same time.
\vskip 2pt
Middle and to right: the representation of the dodecagrid, using projection of the tiles in 
contact with~$H$ onto~$H$. Middle picture: projection of the tiles which stand above~$H$.
Right-hand side picture: projection of the tiles which hang below~$H$.
\end{fig}
\vskip 8pt
}

   In all these figures, we use the projection of a dodecahedron on the plane of its face~0.
It is a central projection from a point which we can imagine on the vertical line which 
passes through the centre of the face, in the direction of face~11 which is opposite to
face~0. This projection was first devised by Schlegel, a geometer of the 19$^{\rm th}$
century: accordingly, the projection is called a Schlegel diagram. The leftmost picture of 
the figure indicates how the faces of a dodecahedron are numbered. The other two pictures
illustrate the representation of the part of the dodecagrid which is used by
the construction of the cellular automaton. The dodecahedra are aligned along the line
which is the intersection of~$H$ and~$V$. The middle picture represents the dodecahedra which
are above~$H$, the right-hand side picture represents those which are below~$H$. In both
cases, each dodecahedron is projected on its face~0 which, by definition is the face in
contact with~$H$. 

   In this representation, the faces which are along an edge belonging to
two neighbouring faces~0 are represented twice as the face they represent belongs to
both dodecahedra which share it. This property is used to indicate the state of a cell
in the pictures of this section. In the pictures, states are represented by colours. 
As a face perpendicular to~$\cal H$ appears twice, it takes the colour of the cell to which
it does not belong. This is why the state of a cell~$c$ appears in the picture as a ring of
faces of the same colour which surround the face~0 of~$c$.

The correspondence between the colours and the states will be given later.

   Now, it is clear that on a tessellation we can construct cellular automata. 
In our constructions, we shall deal with {\bf rotation invariant} cellular automata.
This means that if we perform a rotation
around a cell which leaves the tiling globally invariant, this does not change the 
state of the cell. In the case of the plane, such a rotation is simply a circular
permutation on the numbers of the neighbours of the cell. Checking that a set of rules
is rotation invariant amounts to check that any circular permutation on the neighbours
does not change the new state: the corresponding new rule also belongs to the set.
In the hyperbolic $3D$ space, the characterization is more complex as the group
of the rotations leaving the tiling globally invariant is in fact the group~$A_5$.
However, checking the rotation invariance of a set of rules can be performed by a computer 
program according to a rather simple algorithm, see for instance~\cite{mmbook3}.

    For implementation purposes, it is important to locate the cells as more
conveniently as possible. We refer the reader to~\cite{mmbook1,mmbook2,mmbook3} where
such a convenient system is proposed, based on a tree structure. It is there illustrated 
by many applications, in particular in~\cite{mmbook3}
by the construction of many weakly universal cellular automata. However, for the presentation
of this work, we do not need to introduce this localisation system as the concerned
cells are mainly placed along a line. 

\subsection{Continuing a segment}
\label{continuation}

    In this sub-section, we remind the construction of~\cite{mmstrong} for continuing
a segment along a given line~$\delta$ we call the {\bf guideline}.

    The constraint of the construction lies in the fact that we wish to construct a
{\bf rotation invariant} cellular automaton. From this constraint, the cells do not 
know {\it a priori} their orientation, so that we have to provide them such information. 
In all the tilings we consider, the cells are constructed along a line which does not
cross the cells but which defines their common border. Sub-subsections~\ref{cont_penta}
\ref{cont_hepta} and~\ref{cont_dod} deal with the implementation in the pentagrid, the
heptagrid and the dodecagrid respectively.

\subsubsection{In the pentagrid}
\label{cont_penta}
   
   The guideline~$\delta$~is defined by the common side shared by a yellow cell and
a red one and it is also shared by the two green cells when they appear. 
The guideline defines two half-planes~$\pi_1$ and~$\pi_2$ whose intersection is the
guideline itself. The key of the process is that the guideline is continued by the 
construction of two sequences of cells in parallel. One sequence lies in~$\pi_1$
while the other lies in~$\pi_2$.

    This is illustrated by figures~\ref{propa_penta}.
In this figure, we assume that the initial configuration consists of six cells:
see the leftmost picture in the first row of pictures in Figure~\ref{propa_penta}. 

\def\dR{\hbox{\zz R}}
\def\GG{\hbox{\zz G}}
\def\WW{\hbox{\zz W}}
\def\pY{\hbox{\zz pY}}
\def\YY{\hbox{\zz Y}}
\def\MM{\hbox{\zz M}}  

   The process is the following. We have five states 
denoted by \WW, \YY, \dR, \pY{} and \GG, which we call {\bf blank}, {\bf yellow},
{\bf red}, {\bf pale yellow} and {\bf green} respectively. The blank is the
quiescent state of our automaton. In the figure,
the blank is represented by a blue colour and we use several hues of blue in order to
remember the tree structure of the tiling. 

   The rules are indicated in Table~\ref{tab_propa_penta} after the following format:
\hbox{\tt w$_0$w$_1$w$_2$w$_3$w$_4$w$_5$w$_6$} where {\tt w$_0$} is the current state
of the cell, {\tt w$_i$} is the state of its neighbour~$i$, $i$~in $\{1..5\}$
and {\tt w$_6$} is the new state of the cell after the application of the rule.
The neighbours are increasingly numbered by counter-clockwise turning around the cell.
As the rules are rotation invariant, it is not important which is neighbour~1.

\vtop{
\vskip 5pt
\ligne{\hfill\includegraphics[scale=0.6]{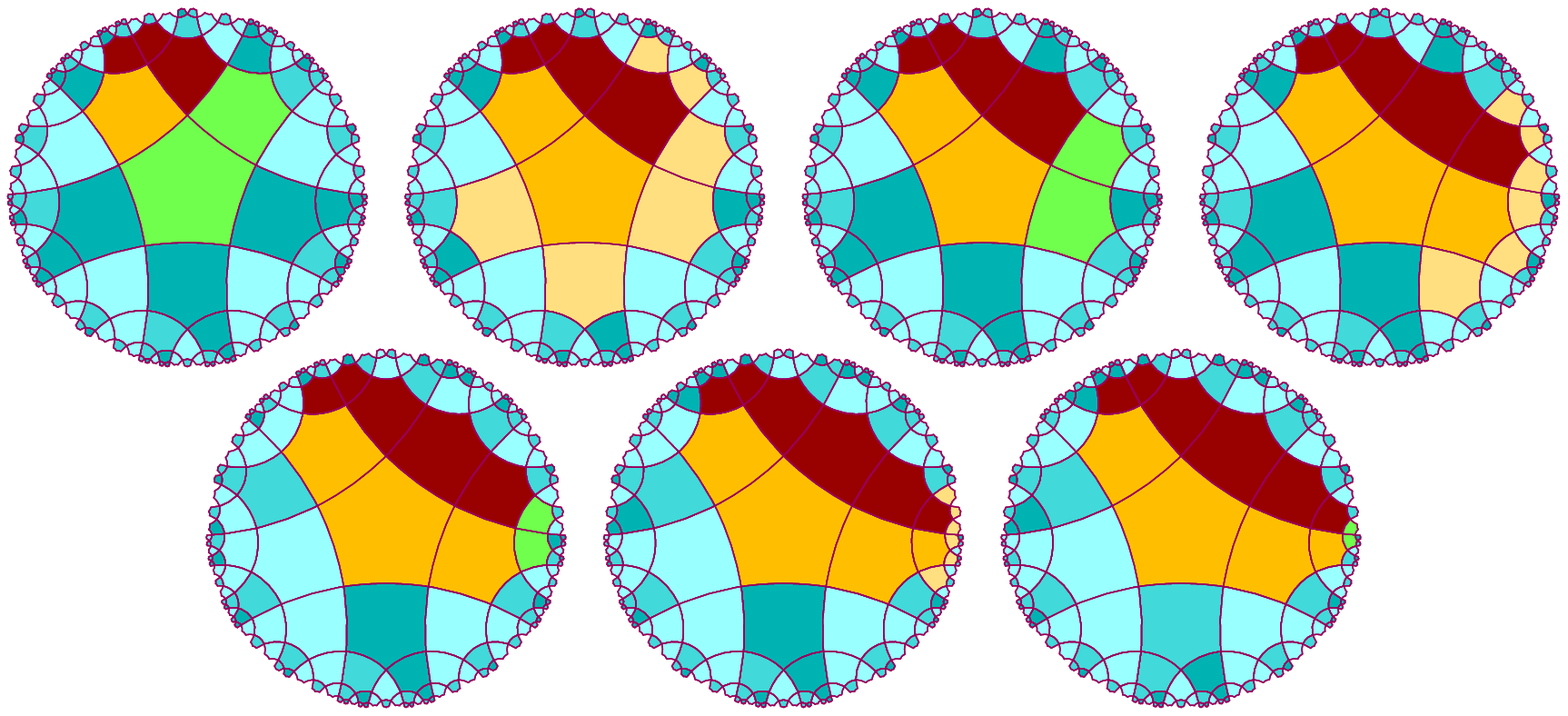}\hfill}
\vspace{-10pt}
\begin{fig}\label{propa_penta}
\leurre
The continuation process in the pentagrid.
\end{fig}
\vskip 8pt
}

   We can formulate the rules as follows. The initial configuration is given
by the first picture. It  consists of three red cells, a yellow
one and two green ones in the configuration indicated by the picture. A blank cell with a 
single green neighbour becomes pale yellow. A green cell with a red, yellow neighbour becomes 
red, yellow respectively. A pale yellow cell with a single non blank
neighbour becomes blank. A pale yellow cell with three blank neighbours and a pale
yellow one becomes green. Yellow and red cells keep their colour.

\vtop{
\begin{tab}\label{tab_propa_penta}
\leurre
Rules for the continuation of the line in the pentagrid.
The rules are rotationally independent.
\end{tab}
\vspace{-12pt}
\grostrait
\setbox110=\vtop{\leftskip 0pt\parindent 0pt\hsize=130pt
{\zz
\obeylines
\leftskip 0pt
\obeyspaces\global\let =\ \parskip=-2pt
--
--     0  1  2  3  4  5  6
--
0 0    W  W  W  W  W  W  W 
  0    W  R  W  W  W  W  W 
  0    W  Y  W  W  W  W  W 
  0    R  Y  R  W  W  W  R 
  0    R  R  R  W  W  W  R 
  0m   R  R  Y  G  W  W  R
  0    Y  G  R  R  W  W  Y
  1    G  R  G  W  W  W  R
  1    G  G  Y  W  W  W  Y
  1    W  G  W  W  W  W pY
\par}
}
\setbox112=\vtop{\leftskip 0pt\parindent 0pt\hsize=130pt
{\zz
\obeylines
\leftskip 0pt
\obeyspaces\global\let =\ \parskip=-2pt
--
--     0  1  2  3  4  5  6
--
1 0    Y  Y  R  R  W  W  Y  
  0    Y  R  Y pY pY pY  Y
  0    R  R  Y  R  W  W  R
  0    R  R  Y pY pY pY  R 
  1   pY  R  W  W  W  W  W 
  1   pY  Y  W  W  W  W  W 
  1   pY  R pY  W  W  W  G 
  1   pY pY  Y  W  W  W  G 
2 0m   Y  G  R  Y  W  W  Y
3 0    Y  Y  R  Y  W  W  Y
\par}
}
\ligne{\hfill\box110 \hfill\box112 \hfill}
\vskip 7pt
\demitrait
\vskip 7pt
}

\subsubsection{In the heptagrid}
\label{cont_hepta}

   Due to the angle~$\displaystyle{{2\pi}\over3}$, the guideline is not defined
by the side of a tile. It is defined by the line which joins the midpoints of consecutive
sides belonging alternately to the same and the consecutive heptagons, 
see~\cite{mmbook1,mmbook3}. Consequently it crosses the tiles of both sequences
whose growth consists in the continuation of the process. We have again five states
and we use the same colours as in the case of the pentagrid to represent the states. 

    This is illustrated by Figure~\ref{propa_hepta}. In this figure, we assume
that the initial configuration consists of five cells: see the leftmost picture
of the first row in Figure~\ref{propa_hepta}. Although each rule is longer than those
for the pentagrid, we shall see that the rules for the heptagrid are simpler than
those of Sub-subsection~\ref{cont_penta}.

   The format of the rules is alike that for the pentagrid: here, each cell has simply
seven neighbours instead of five of them. Hence the new state taken by the cell
after the rule is applied is {\tt w$_8$} in Table~\ref{tab_propa_hepta}.

\vtop{
\vskip 5pt
\ligne{\hfill\includegraphics[scale=0.6]{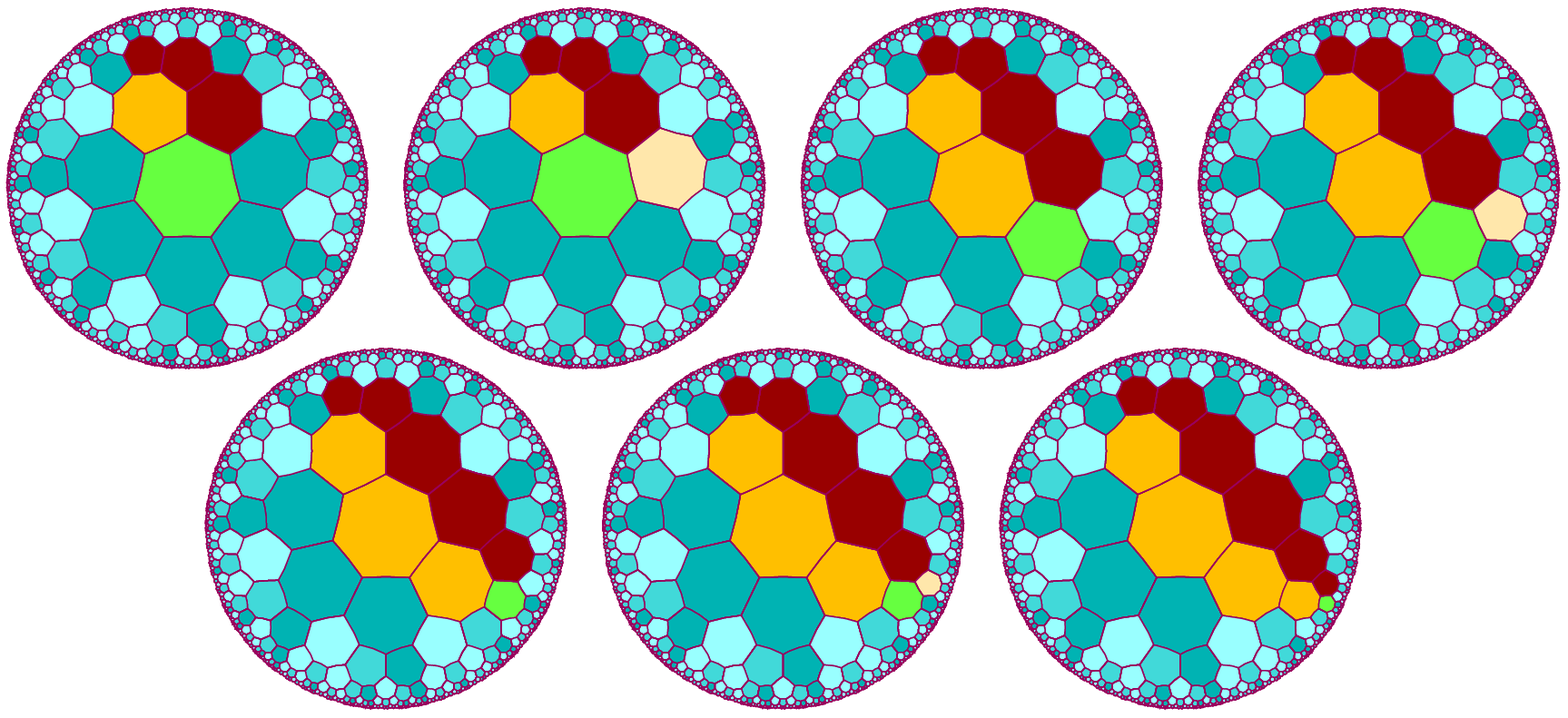}\hfill}
\vspace{-10pt}
\begin{fig}\label{propa_hepta}
\leurre
The continuation process in the heptagrid.
\end{fig}
\vskip 8pt
}

However the rules are a bit simpler as can be seen on the following informal
presentation:

    A white cell with a red and a green neighbour around the same vertex,
the red cell first while counter-clockwise turning around the cell, becomes pale yellow.
The green cell remains green if it has no pale yellow neighbour. If it has a pale
yellow neighbour, it becomes yellow. A pale yellow cell becomes red at the next
time. A blank cell with a pale yellow and a green neighbour, in this order, sharing
a common vertex becomes green. Yellow and red cells keep their colour.

\vtop{
\begin{tab}\label{tab_propa_hepta}
\leurre
Rules for the continuation of the line in the heptagrid. 
The rules are rotationally independent.
\end{tab}
\vspace{-12pt}
\grostrait
\setbox110=\vtop{\leftskip 0pt\parindent 0pt\hsize=154pt
{\zz
\obeylines
\leftskip 0pt
\obeyspaces\global\let =\ \parskip=-2pt
--
--     0  1  2  3  4  5  6  7  8
--
0 0    W  W  W  W  W  W  W  W  W
  0    W  Y  W  W  W  W  W  W  W
  0    W  R  W  W  W  W  W  W  W
  0    W  G  W  W  W  W  W  W  W
  0    W  Y  R  W  W  W  W  W  W
  0    W  R  R  W  W  W  W  W  W
  0    W  G  Y  W  W  W  W  W  W
  0    R  Y  R  W  W  W  W  W  R
  0    R  R  Y  R  W  W  W  W  R
  0    R  R  Y  G  W  W  W  W  R
  0    Y  G  R  R  R  W  W  W  Y
  0    G  R  Y  W  W  W  W  W  G 
  1    W  R  G  W  W  W  W  W pY 
\par}
}
\setbox112=\vtop{\leftskip 0pt\parindent 0pt\hsize=154pt
{\zz
\obeylines
\leftskip 0pt
\obeyspaces\global\let =\ \parskip=-2pt
--
--     0  1  2  3  4  5  6  7  8
--
1 0    R  R  Y  G pY  W  W  W  R
  0    W pY  W  W  W  W  W  W  W
  0    W  R pY  W  W  W  W  W  W
  0    Y  Y  R  R  R  W  W  W  Y
  1   pY  R  G  W  W  W  W  W  R
  1    G pY  R  Y  W  W  W  W  Y 
  1    W pY G   W  W  W  W  W  G
2 0m   Y  G  R  R  Y  W  W  W  Y 
  0    W  Y  Y  W  W  W  W  W  W
  0    R  R  Y  Y  R  W  W  W  R 
4 0    Y  Y  R  R  Y  W  W  W  Y
\par}
}
\ligne{\hfill\box110 \hfill\box112 \hfill}
\vskip 7pt
\demitrait
\vskip 7pt
}

\subsubsection{In the dodecagrid}
\label{cont_dod}

    Here, the guideline is again a line which does not cross the tiles: it is the 
intersection of the planes~$\cal H$ and~$\cal V$.

    Basically, the process looks like that of the pentagrid. However, we have a
small adaptation for the $3D$~situation. The guideline is shared by four tiles 
while, in the case of the pentagrid it is shared by two of them only. This means that
here, the continuation requires the construction in parallel of four sequences
of tiles. The planes~$\cal H$ and~$\cal V$ define four regions $\eta_1$, $\eta_2$,
$\eta_3$ and~$\eta_4$ of the hyperbolic
$3D$~space. Each of them is the half of a half-space. Each sequence of tiles lies in its own
region~$\eta_i$, different from the regions of the others. Moreover, we distinguish
a upper part and a lower with respect to~$\cal H$: two regions are above~$\cal H$
while the two others are below it. 

   We use again the colours defined for the pentagrid. One sequence of the upper ones will 
be the yellow one, the other will be mauve, a new colour, denote by~\MM. Both sequences
below~$\cal H$ will be red. Accordingly, each sequence knows whether it is above or
below~$\cal H$. But we shall see that this may sometimes be disturbed.  

\vtop{
\ligne{\hfill
\includegraphics[scale=0.6]{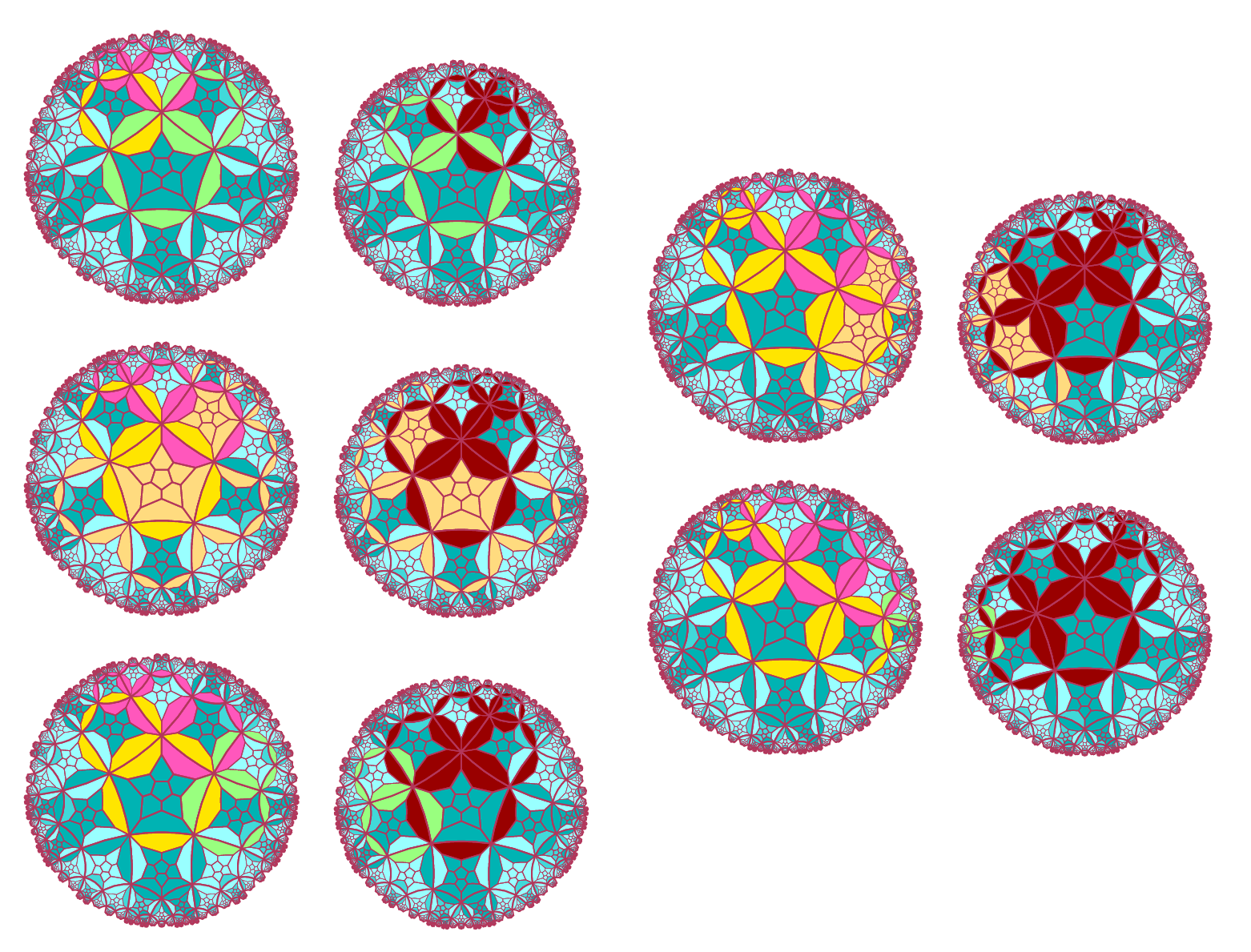}
\hfill
}
\vspace{-5pt}
\begin{fig}
\label{propdod}
\leurre
The continuation process in the dodecagrid. In each pair, the left-, right hand side picture
is the projection of the cells above, below~$\cal H$ respectively.
\end{fig}
}

   Figure~\ref{propdod} illustrates the process, while Table~\ref{propa_dodec} gives
the rules which control the process.

   Initially, there are twelve cells: six of them above~$\cal H$ and six of them
below~$\cal H$. The yellow cell, three mauve ones and two green ones which are
above~$\cal H$ reproduce the initial configuration for the pentagrid, see
Figure~\ref{propa_penta}. Below~$\cal H$, the four red cells are below the mauve and yellow ones.
The two green cells are below those which are above~$\cal H$.

   The format of the rules is a bit different from what we have seen in the planar tilings.
Here, a rule is written as 
\hbox{\tt w$_o$w$_0$w$_1$w$_2$w$_3$w$_4$w$_5$w$_6$w$_7$w$_8$w$_9$w$_{10}$w$_{11}$w$_n$},
where {\tt w$_o$} is the current state of the cell, {\tt w$_i$} with $i$~in $\{0..11\}$
the state of the neighbour~$i$, {\it i.e.} the one which seen through the face~$i$
of the cell. The new state of the cell after the rule is applied is {\tt w$_n$}.

   The rules can be formulated like this.
A blank cell which sees a green one, state~{\tt dG},
becomes pink, state~{\tt dY}. A pink cell becomes blank unless it has two pink neighbours 
through faces sharing an edge: one pink neighbour is above~$\cal H$ and the other is below.
In that case, the cell becomes green. At last, the green cell takes the colour of its neighbour
on the same line: in particular, if the cell is on the yellow line, it takes the blank
of~\LN$_9$. Mauve, yellow and red cells keep their colour.
Figure~\ref{propa_dodec} illustrates this process giving, at each time two projections
onto~$\cal H$: on the left-hand side, the cells above the plane, on the right-hand side,
the cells below it.

\setbox114=\vtop{\leftskip 0pt\parindent 0pt\hsize=265pt
{\zz
\obeylines
\leftskip 0pt
\obeyspaces\global\let =\ \parskip=-2pt
--
--   old  0  1  2  3  4  5  6  7  8  9 10 11  new
--
0 0    W  W  W  W  W  W  W  W  W  W  W  W  W  W
  0    W  M  W  W  W  W  W  W  W  W  W  W  W  W
  0    W  Y  W  W  W  W  W  W  W  W  W  W  W  W
  0    W  R  W  W  W  W  W  W  W  W  W  W  W  W
  0    M  R  M  Y  W  W  W  W  W  W  W  W  W  M
  0    M  R  M  M  W  W  W  W  W  W  W  W  W  M
  0m   M  R  G  Y  M  W  W  W  W  W  W  W  W  M
  0    Y  R  M  M  G  W  W  W  W  W  W  W  W  Y
  0    R  M  R  R  W  W  W  W  W  W  W  W  W  R 
  0    R  Y  R  R  W  W  W  W  W  W  W  W  W  R 
  0    R  M  R  R  G  W  W  W  W  W  W  W  W  R 
  0    R  Y  G  R  R  W  W  W  W  W  W  W  W  R 
  0    W  G  W  W  W  W  W  W  W  W  W  W  W pY
  1    G  G  Y  G  W  W  W  W  W  W  W  W  W  Y 
  1    G  G  G  M  W  W  W  W  W  W  W  W  W  M
  1    G  G  G  R  W  W  W  W  W  W  W  W  W  R
  1    G  G  R  G  W  W  W  W  W  W  W  W  W  R
1 0    M  R  M  Y  M  W  W  W  W  W  W  W  W  M   
  0    Y  R  M  M  Y  W  W  W  W  W  W  W  W  M
  0    M  R  Y  M pY pY pY pY pY pY pY pY pY  M
  0    Y  R  Y  M pY pY pY pY pY pY pY pY pY  Y
  0    R  M  R  R  R  W  W  W  W  W  W  W  W  R    
  0    R  Y  R  R  R  W  W  W  W  W  W  W  W  R    
  0m   R  M  R  R pY pY pY pY pY pY pY pY pY  R  
  0m   R  Y  R  R pY pY pY pY pY pY pY pY pY  R  
  1   pY  M  W  W  W  W  W  W  W  W  W  W  W  W
  1   pY  Y  W  W  W  W  W  W  W  W  W  W  W  W
  1   pY  R  W  W  W  W  W  W  W  W  W  W  W  W
  1   pY pY  M  W  W  W  W  W  W  W  W  W  W  W
  1   pY pY  Y  W  W  W  W  W  W  W  W  W  W  W
  1   pY pY  Y pY  W  W  W  W  W  W  W  W  W  G
  1   pY pY pY  M  W  W  W  W  W  W  W  W  W  G
  1   pY pY pY  R  W  W  W  W  W  W  W  W  W  G
  1   pY pY  R pY  W  W  W  W  W  W  W  W  W  G
  0m   W pY pY  W  W  W  W  W  W  W  W  W  W  W
2 0m   Y  R  Y  M  G  W  W  W  W  W  W  W  W  Y 
3 0    Y  R  Y  M  Y  W  W  W  W  W  W  W  W  Y 
--
\par}
}
\vtop{
\begin{tab}\label{propa_dodec}
\leurre
Rules for the propagation of the $1D$-structure in the dodecagrid.\vskip 0pt
\noindent
The rules are rotationally independent.
\end{tab}
\vspace{-12pt}
\grostrait
\ligne{\hfill\box114 \hfill}
\vskip 7pt
\demitrait
\vskip 7pt
}

   In the next subsection, the automaton defined by these rules will be denoted
by~$\cal P$. Note that $\cal P$ has five states in the case of the planar tilings and
it has six states in the case of the dodecagrid.

\subsection{The results and their proofs}
\label{stronghyp}

    As already mentioned, we take into consideration \LN$_9$ only. We are now with a cellular 
automaton with 9~states which is actually universal up to the halting of its own computation. 
The automaton has 5 states or~6 states depending on whether we consider the planar tilings
or the dodecagrid respectively. This means that the simple superposition of~\LN$_9$ with~$\cal P$
would give us~14 or 15 states respectively.

   Following the idea given in~\cite{mmstrong}, we can improve such a result by
identifying states of~$\cal P$ with those of~\LN$_9$. In~\cite{mmstrong}, the identifications
where performed according to the scheme~$(S_1)$. In this scheme, the first line indicates
the states of~\LN$_9$, the second and the third ones those of~$\cal P$ for the planar tilings 
and the dodecagrid respectively. The identified states are placed one upon another. Of course,
\YY{} is identified with any state of~\LN$_9$. This means that, a priori, several states
have a double interpretation. As an example, {\tt A} is a state of~\LN$_9$ while, by the 
identification, it is also the pink state in~$\cal P$. The distinction is obtained by the
neighbouring of the considered cell. To take this example, a pink cell has a lot of
{\tt W}-neighbours, more than a yellow cell. This entails that a cell containing~{\tt A} knows
which rules it has to apply: those of~\LN$_9$ or those of~$\cal P$.
   
\newdimen\largesub\largesub=20pt
\def\subst #1 #2 #3 #4 #5 #6 #7 #8 #9 {%
\zz
\hbox to \largesub{\hskip 5pt#1\hfill}
\hbox to \largesub{\hskip 5pt#2\hfill}
\hbox to \largesub{\hskip 5pt#3\hfill}
\hbox to \largesub{\hskip 5pt#4\hfill}
\hbox to \largesub{\hskip 5pt#5\hfill}
\hbox to \largesub{\hskip 5pt#6\hfill}
\hbox to \largesub{\hskip 5pt#7\hfill}
\hbox to \largesub{\hskip 5pt#8\hfill}
\hbox to \largesub{\hskip 5pt#9\hfill}
}
\def\substb #1 #2 {%
\zz
\hbox to \largesub{\hskip 5pt#1\hfill}
\hbox to \largesub{\hskip 5pt#2\hfill}
}
\vskip 5pt
\setbox110=\vtop{\leftskip 0pt\parindent 0pt\hsize=284pt
\ligne{\subst {\theblank} {0} {1} {y} {A} {B} {T} {U} {4} \hfill}
\ligne{\subst {Y} {Y} {Y} {Y} {\hskip-4pt pY} {Y} {R} {} {} 
\substb {W} {G} \hfill}
\ligne{\subst {Y} {Y} {Y} {Y} {\hskip-4pt pY} {M} {R} {} {} 
\substb {W} {G} \hfill}
}
\ligne{\hskip 60pt
$\vcenter{\box110}$\hfill$(S_1)$\hskip 20pt}
\vskip 7pt

    The situation may be more intricate with other states, but it is not very difficult,
we refer the reader to~\cite{mmstrong} for corresponding explanations. There is just a 
difference with~\cite{mmstrong} where there was no process of continuation in the $1D$-cellular
automaton. Here we have two continuation processes. The first one concerns the construction
of the hyperbolic structure. We can notice that the speed of the growth of this structure
is~$\displaystyle{1\over2}$, as a new cell is appended each second time. Now, in the yellow 
cells, we also have the continuation process of~\LN$_9$ which also advances at the
speed~$\displaystyle{1\over2}$: the distance between both processes remains constant
as long as the stopping signal did not reach the continuation process of~\LN$_9$. But
this raises no difficulty and can easily be absorbed by the scenario depicted 
in~\cite{mmstrong}.

    This allows us to state:

\begin{thm}\label{stronghyp11}
In each of the following tilings: the pentagrid and the heptagrid of the hyperbolic
plane, there is a deterministic, rotation invariant cellular 
automaton with radius~$1$ 
which has $11$~states and which is 
strongly universal.
\end{thm}

   The goal of this section is to prove the following result:

\begin{thm}\label{stronghyp10}
In each of the following tilings: the pentagrid and 
the heptagrid of the hyperbolic
plane and also 
the dodecagrid of the hyperbolic $3D$-space, there is a deterministic, 
rotation invariant cellular automaton with radius~$1$ which has $10$~states and which is 
strongly universal.
\end{thm}

   Of course, Theorem~\ref{stronghyp11} is also a corollary of Theorem~\ref{stronghyp10},
and as such, it also holds for the dodecagrid. However, following~\cite{mmstrong}, it
can be proved for the pentagrid and the heptagrid, without reference to
Theorem~\ref{stronghyp10}.

   The identification which we shall now prove is given by~$(S_2)$.
\vskip 5pt
\setbox110=\vtop{\leftskip 0pt\parindent 0pt\hsize=284pt
\ligne{\subst {\theblank} {0} {1} {y} {A} {B} {T} {U} {4} \hfill}
\ligne{\subst {W} {Y} {Y} {Y} {\hskip-4pt pY} {Y} {R} {} {} 
\substb {G} {} \hfill}
\ligne{\subst {W} {Y} {Y} {Y} {\hskip-4pt pY} {M} {R} {} {} 
\substb {G} {} \hfill}
}
\ligne{\hskip 60pt
$\vcenter{\box110}$\hfill$(S_2)$\hskip 20pt}
\vskip 7pt

   In Sub-subsection~\ref{ident} we explain why the identification is possible,
and in Sub-subsection~\ref{rules10}, we give the rules of the automaton proving
Theorem~\ref{stronghyp10} after the identification.

   Before turning to this problem, let us see why we cannot go further than~$(S_2)$
in the identifying process. 

   In~\cite{mmstrong} and in the construction we perform to prove Theorem~\ref{stronghyp10},
the stopping signal is a state of~$\cal P$ which is captured by the red line, the mauve line
in the case of the dodecagrid, and which travels to the right-hand side of the configuration.
During this motion, the signal destroys all states which do not belong to the Turing machine,
including the constructing pattern of the lines when it reaches the pattern. In~\cite{mmstrong}
and here, the stopping signal is identified to~{\tt G}. Due to its nature, the stopping
signal must not be seen on the Turing part of the line before the end of the 
Turing computation and it also must not be seen until this end by an ordinary red or mauve cell, 
otherwise, the end of the computation would occur too soon, ruining the simulation.
 
As shall soon be seen, {\tt W} is identified with the
blank of~\LN$_9$. This identification is not trivial and, except the case of the
heptagrid, it requires to modify $\cal P$ for the pentagrid and for the dodecagrid.
Now, it would be necessary to identify {\tt G} with a state of~\LN$_9$ too.
And this is not possible: besides~{\tt T} and~{\tt 4}, all states of~\LN$_9$
occur many times during the computation and are repeated in many places seen by ordinary 
red or mauve cells. For this reason, as it is seen from the beginning by ordinary red or
mauve cells, {\tt 4} cannot be the stopping signal. For what is {\tt T}, it would occur
in between two blanks of~\LN$_9$. This means that two heads would be raised, starting
new computations which are not at all connected with the simulation. And so,
{\tt G} remains the single candidate for the stopping signal and from what we have seen,
it cannot be identified with a state of~\LN$_9$. We denote by~$\cal Q$ the new automaton
resulting from this identification and applying the rules of~\LN$_9$ to the cells of the yellow
line and the rules of~$\cal P$ for the cells of another type.

\subsubsection{Proof of that the identification~$(S_2)$ holds}
\label{ident}

   We successively examine the situation in the pentagrid, in the heptagrid and in the
dodecagrid. The initial configuration is similar in all cases and it will be precisely
described in the case of the pentagrid only.

\vskip 5pt
\ligne{\hskip-20pt$\vcenter{\underline{\hbox{Pentagrid}}}$
\hfill}
\vskip 5pt

   In this case, the identification of~\theblank{} with~\WW{} raises
a contradiction between the rules of~\LN$_9$ and those of~$\cal P$. During the computation,
many Turing symbols remain some time in between two blanks. This is the case for~{\tt A}
for instance. Now, such a state must not be changed, as long as it is not involved in
a collision with an impulsion. Now, according to the identification, the neighbourhood
of such~{\tt A} in the scenario depicted in Sub-subsection~\ref{cont_penta}
is \hbox{\zz T W W W W}, as it can see only a red cell as a non-blank one. The rule
then should be \hbox{\zz A T W W W W A}. This is in contradiction
with the rule \hbox{\zz A T W W W W W} telling that a pale yellow cell seeing
a single non-blank cell becomes blank.

   Accordingly, the process of Sub-subsection~\ref{cont_penta} must be tuned.

Necessarily, we have to keep the rule \hbox{\zz A T W W W W W}. The solution is to reinforce
the way which allows a cell to know for sure that it is on the yellow line. For that
purpose, we introduce a new pattern for the yellow cell. We decide that such a cell have a 
neighbourhood of the form \hbox{\zz T l B B r}, where {\tt r}, {\tt l} is the state of the 
left-, right-hand side neighbour respectively of the considered yellow cell~$c$ whose
state we denote by~{\tt c}.

   Figure~\ref{new_propa_penta} illustrates the new implementation of the whole computation
based on this new setting for the yellow cells. The three rows of the figure illustrate the 
three steps of the computation.

   During the first step, $\cal Q$ performs the computation of the simulated Turing machine
and $\cal P$ constructs the line needed for the computation of~\LN$_9$. At this point,
we can take advantage of~$\cal P$ in order to provide the needed background to the yellow line.
For this purpose, we no more need the action between the states~{\tt U} and~{\tt B} as
depicted in Section~\ref{implement}. It is enough to decide that the yellow cell produced
by the transformation of the green cell of the yellow line is the state~{\tt U}. Then,
the rules of~\LN$_9$ with the states~{\tt U} and~{\tt 4} allow to create the needed
two blanks between two occurrences of~{\tt 0} each time the head of the Turing machine
goes out outside the current configuration on its right-hand side end. Accordingly,
this will slightly simplify the action of the stopping signal. 
Also, in the initial configuration, the segment $[c,d]$ contains the initial configuration
of~\LN$_9$. Giving 1 as coordinate for the first yellow cell at the left-hand side of the
configuration, $c$~can be given the coordinate~5, cell~1 being in~{\tt B}, the cells~2 up
to~4 being blank. We decide
that~$d$ contains the last Turing symbol. Then,starting from \hbox{$d$+2} until
\hbox{$d$+4} the cells contain~{\tt U} and the cell \hbox{$d$+5} is green.

   The last two rows of Figure~\ref{new_propa_penta} are devoted to the propagation
of the stopping signal and its action. 
   
   The second row of the figure shows us the occurrence of the stopping signal on the yellow
line, its immediate transfer to the red line and there, its progression towards the
right-hand side end of the configuration. The last row shows us what happens when
the stopping signal arrives very close to the end of the configuration and how it stops
its progression.

   Provided that rules can be defined to perform these various actions, we shall look at this
point in Sub-subsection~\ref{rules10}, we can now look at the different neighbourhood of these
cells in order to check that the pattern we devised for the yellow cells also avoids
any confusion in the computation of~$\cal P$. As we identified the states of~$\cal P$ with
those of~\LN$_9$, {\tt G} being excepted, we shall formulate the new rules for $\cal P$
in terms of states of~\LN$_9$ and~{\tt G} when this latter state is involved.

\vtop{
\vskip 5pt
\ligne{\hfill\includegraphics[scale=0.5]{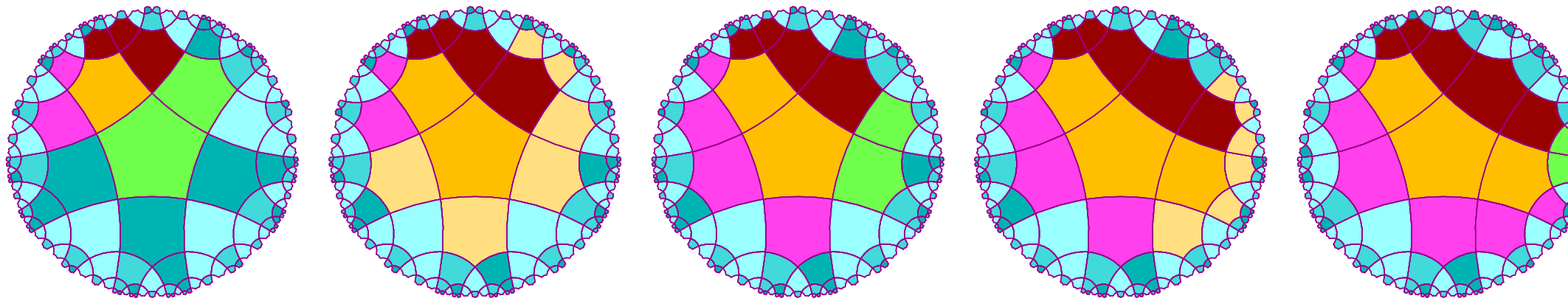}\hfill}
\ligne{\hfill\includegraphics[scale=0.5]{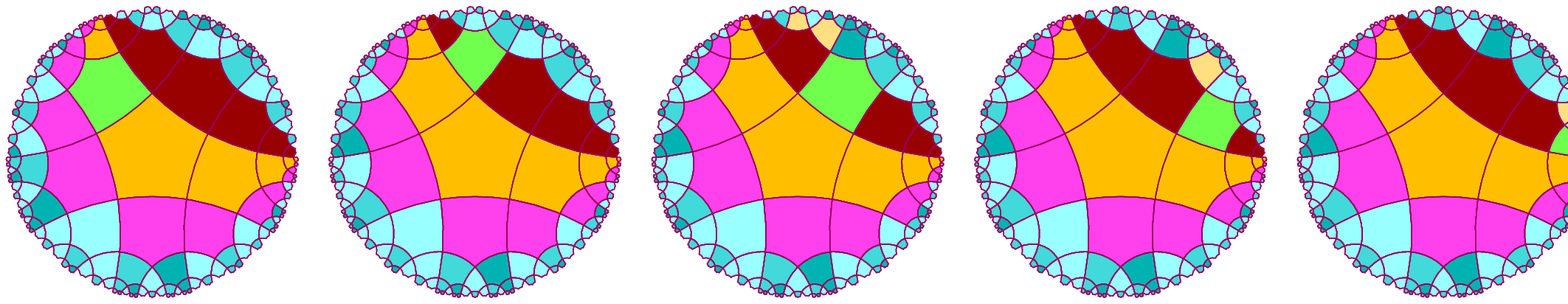}\hfill}
\ligne{\hfill\includegraphics[scale=0.5]{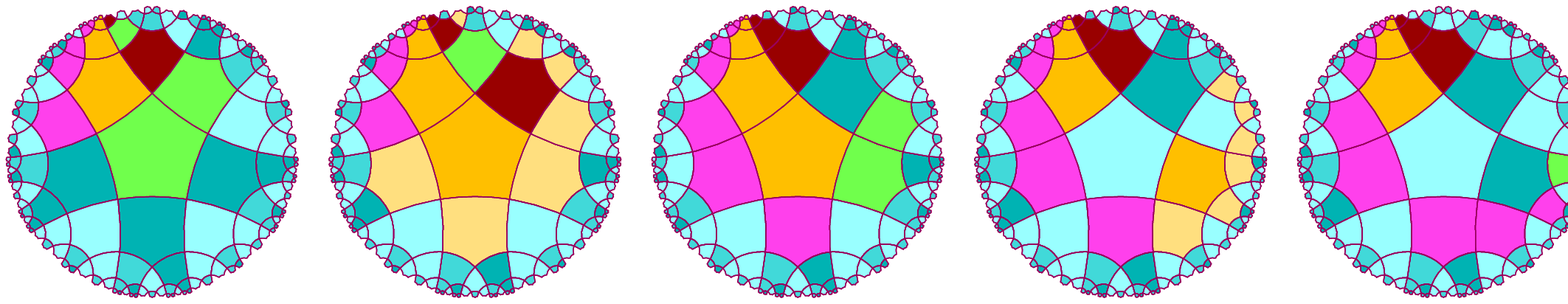}\hfill}
\vskip 10pt
\begin{fig}\label{new_propa_penta}
\leurre
The new continuation in the pentagrid. 
First line: before the stopping signal. Second line: the stopping signal was just triggered.
Third line: the stopping signal reaches the pattern which constructs the line.
\end{fig}
\vskip 8pt
}

   Let us first look at the yellow line. 

   We decide that the neighbourhood of a yellow cell is of the form
\hbox{\tt TlBBr}, where {\tt l}, {\tt r} is the state of the left-, right-hand side neighbour
of the cell on the yellow line. We have to check that this allows a yellow cell to know that 
it is such a cell and to correctly identify its left- and right-hand side neighbours on the 
yellow line.

   Indeed, if {\tt l} or {\tt r} is neither~{\tt B} nor~{\tt T}, denote by~{\tt a} the neighbour
which is different from both~{\tt B} and~{\tt T}. Then, necessarily, {\tt a} is in between
{\tt B} and the {\tt T} which belongs to the red line. This allows to identify correctly
whether {\tt a} is~{\tt l} or{\tt r}. Now, consider the case when both {\tt l} and~{\tt r}
are equal to~{\tt B} or~{\tt T}. They cannot be both equal to~{\tt T} as there is a 
single~{\tt T} on the yellow line. The can be both equal to~{\tt B}, in which case 
\hbox{\tt c $=$ 1}. But this situation gives rise to a single rule. And so, we consider the
case when \hbox{\tt $\{$l$,$r$\}$ = $\{$B$,$T$\}$}. Depending on which {\tt T} is considered
as that of the red line we have to consider the neighbourhood \hbox{\tt BcT} or 
\hbox{\tt TcB}. Now, Table~\ref{table_reg} indicates that {\tt c} cannot be in
\hbox{\tt \theblank$,$ B$,$ U$,$ 4}. When {\tt c} is {\tt T} or {\tt 0}, only the
neighbourhood \hbox{\tt BcT} exists on the yellow line which fixes the correct orientation
and a single rule can be applied. When {\tt c} is {\tt y} or {\tt A}, the only
neighbourhood is \hbox{\tt TcB}, also giving rise to a single rule. The remaining case
is \hbox{\tt c $=$ 1} but the two possible rules give the same state. And so, whatever the choice
of the red~{\tt T}, the correct new state will be defined.

  We remain with the other situations regarding the red and yellow line when the computation
of the simulated machine has been completed and what happens at the right-hand side end of 
the configuration. We have tow other states: the detection of the stopping signal with its
progression and the halting of the computation, when the stopping signal reached the end of
the configuration. Tables~\ref{new_rules_penta_yellow}
and~\ref{new_rules_penta_red} show the rules in a way where these contexts are clearly
separated both for the yellow and the for the red cells. A close examination of the tables
show that the neighbourhoods involved in these different kinds of cells can never be confused.

   And so, we can conclude that the new automaton satisfies the requirement of the statement
of Theorem~\ref{stronghyp10} for what is the pentagrid. Accordingly, the theorem
is proved in this case.


\vskip 5pt
\ligne{\hskip-20pt$\vcenter{\underline{\hbox{Heptagrid}}}$
\hfill}
\vskip 5pt

   In the case of the heptagrid, the same problem as in the case of the pentagrid happens
with the state~{\tt A} when the blank of~\LN$_9$ is identified with that of~$\cal P$.
Indeed, the rule \hbox{\zz A T G W W W W W T} may be raised in the yellow line, far from the
end. This would be the case for a yellow state under the state~{\tt A} when the stopping signal
running over the red line arrives in front of cell in~{\tt U} on the yellow line. For that
latter cell, we need the rule \hbox{\zz A T G W W W W W A}.

\vtop{
\vskip 5pt
\ligne{\hfill\includegraphics[scale=0.5]{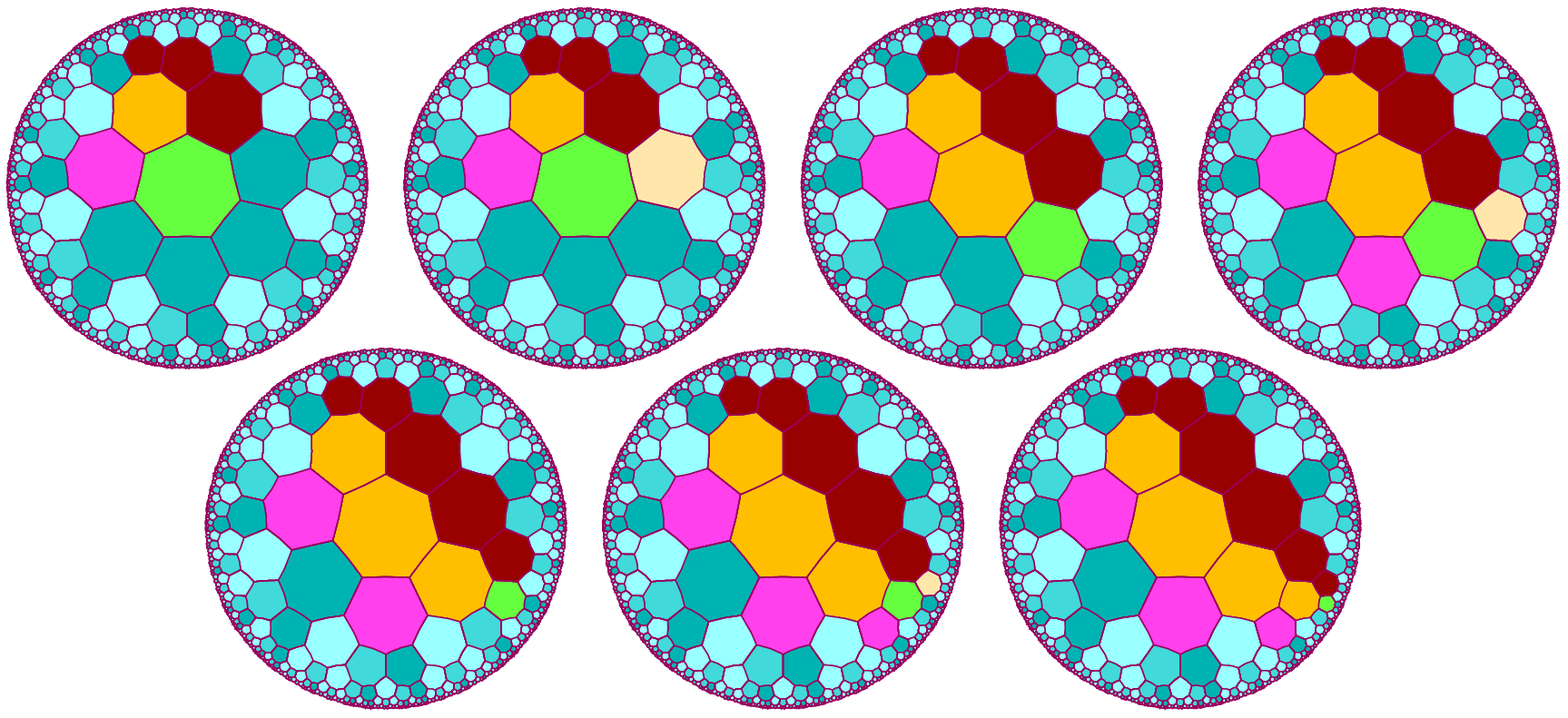}\hfill}
\ligne{\hfill\includegraphics[scale=0.5]{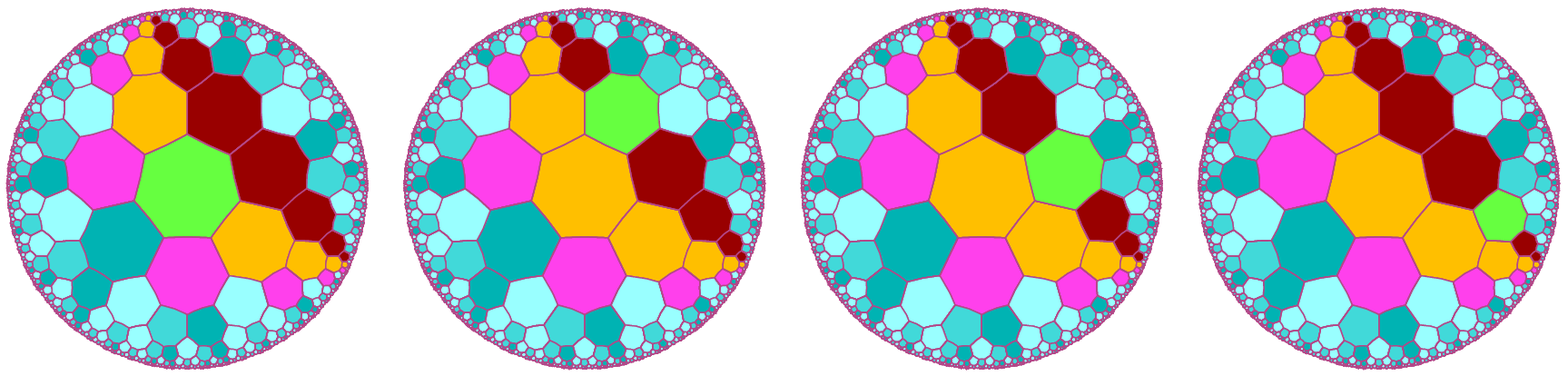}\hfill}
\ligne{\hfill\includegraphics[scale=0.5]{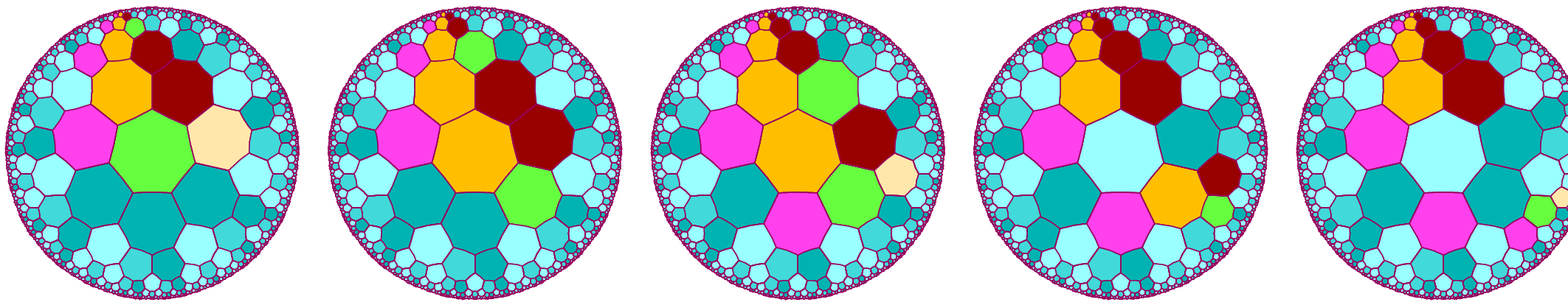}\hfill}
\vskip 10pt
\begin{fig}\label{new_propa_hepta}
\leurre
The new continuation in the heptagrid. 
First line: before the stopping signal. Second line: the stopping signal was just triggered.
Third line: the stopping signal reaches the pattern which constructs the line.
\end{fig}
\vskip 8pt
}

   As in the case of the pentagrid, the solution is to reinforce the identification of the
yellow cells. We decide that they have a neighbourhood of the form
\hbox{\zz T Y B W B Y T}. These two~{\tt  B}'s are present in the initial configuration 
and the construction of the line appends the new ones as indicated in the first two rows
of Figure~\ref{new_propa_hepta}. In the situation of a cell in the state~{\tt A},
we now have two different rules : \hbox{\zz A T G W W W W W T} and
\hbox{\zz A G W B W B W T A}. The new structure of the rules on the yellow line
can also be seen in the rules of the cellular automaton
for the heptagrid as displayed by Table~\ref{new_rules_hepta_whole}. We can check on the rules
that the cells know their position without ambiguity. The pattern~\hbox{\tt BWB} allows a 
more clear identification that the pattern~\hbox{\tt BB} in the case of the pentagrid.
This can clearly be seen from the neighbourhood \hbox{\zz T Y B W B Y T} which we can
rewrite as \hbox{\zz T x B W B z T}. It is plain that whatever the values of~{\tt x}
and~{\tt z} among the states of~\LN$_9$, as {\tt B} never occurs on the red line, the pattern
{\tt BWB} cannot be placed at another site in the word.

\vskip 5pt
\ligne{\hskip-20pt$\vcenter{\underline{\hbox{Dodecagrid}}}$
\hfill}
\vskip 5pt

    In the dodecagrid, the neighbourhood of a yellow cell is \hbox{\zz T B z W W x W W W W W W}
after the identification~$(S_2)$. If \hbox{\tt x $=$ B} and \hbox{\tt z $=$ W},
the cell knows that its back is the face in~{\tt T}, but it may hesitate on which ~{\tt U}
 is its face~1. If it takes one choice, it considers that it has to apply the instruction 
raised by~\hbox{\tt B y W} and if it takes the other~{\tt U}, the instruction is
raised by~\hbox{\tt W y B}. However, Table~\ref{table_reg} shows the rules 
\hbox{\zz B0\theblank $\rightarrow$ y} and 
\hbox{\zz \theblank{}0B $\rightarrow$ 1} and a lot of other examples which are kept by \LN$_9$.

   This forces us to define a new pattern which, as previously, reinforce the identification
of the yellow cells. But this time, the new scenario will reinforce the identification of
each line. The new pattern is illustrated by Figure~\ref{newpattern} and it consists in
putting a non-blank cell on the faces~6 and~7 of the cells, see 
Figure~\ref{propdod} in Subsection~\ref{continuation}.

\vtop{
\vspace{-10pt}
\ligne{\hfill\includegraphics[scale=0.9]{newpattern.ps}\hfill}
\vspace{-20pt}
\begin{fig}\label{newpattern}
\leurre
Note that now, faces~$6$ and~$7$ are non-blank.
\end{fig}
}

As far as faces~6 and~7 are no more blank, there is a single blank face, face~9, which is 
surrounded by blank faces of the dodecahedron. Face~1 is opposite to face~9, so that face~1 is 
clearly identified. Now, we decide that in the case of yellow cells, faces~6 and~7 are 
in~{\tt B} which is identified with the mauve colour. Consider again a cell of the yellow 
line. There is at least one face in~{\tt T} and two contiguous faces in~{\tt B}, not in contact 
with the face in~{\tt T} and these three faces are around face~1 which is immediately identified. 
Let us check that the cell can now identify which face is face~0. If there
is no neighbour of the cell in~{\tt T} on the yellow line, then the cell has a single
neighbour in~{\tt T} and it is seen through face~0 necessarily. Assume that there
is another cell in~{\tt T} and there is necessarily at most a single one. There cannot be 
another one as there is at most one cell in~{\tt T} on the yellow line. Necessarily, 
this {\tt T}-face is in contact with face~0 and with the faces in~{\tt B} around face~1. 
There could be a confusion if the fifth face around face~1 would also be in~{\tt B}. But the
confusion cannot happen. Looking at the patterns {\zz BxT} and {\zz TxB} in the computation 
of~\LN$_9$, see Table~\ref{table_reg}, there is no such patterns when 
\hbox{\tt x $\in$ $\{\theblank,\hbox{\tt B$,$4$,$U}\}$}, one of them exist but not the others
when \hbox{\tt x $\in$ $\{\hbox{\tt 0$,$y$,$A$,$T}\}$}. The single case when both patterns
exist is when \hbox{\tt x $=$ 1} and, in both cases, the corresponding rule produces the
same state. And so, in all cases, but one only, face~0 is clearly identified and in all cases,
the applied rule always yields the right result. 

   Let us decide that faces~6 and~7 are red on the mauve line and on the red lines too.
This allows the cells of these lines to easily know to which line they belong too and
which is their face~0. Remember that a cell of the yellow line has at most 
two {\tt T}-neighbours after the identification of~$(S_2)$: the face which looks at the
red line below~$\cal H$ and possibly, either face~2 or face~5 which are shared by its neighbours
on the yellow line. A cell of the mauve line has three or four {\tt T}-neighbours exactly: 
face~0, face~6 and face~7 and, possibly, its face~1. As its faces~2 and~5 and the cell
itself are mauve or at most one of them is green, this allows the cell to recognize
that it is mauve. A cell of the red line has at least five {\tt T}-neighbours: its
faces~1, 2, 5, 6 and~7. It is not important for a cell of this line to identify
its face~0: the cell is always red.

    Note that this argument has to be tuned for the cells which are at the ends of the 
configuration. At the fixed end, the face~2 or~5, depending on the side with respect 
to~$\cal V$, are~{\tt W}. As these cells belong to the initial configuration, faces~6 and~7 
are fitted with the appropriate colour. This does not change the distinction we have noted,
so that each cell knows to which kind it belongs. Consider the other end where we have the
constructing pattern. Initially, the four cells are green. They have nine {\tt W}-neighbours,
two green neighbours and the last one has the colour of the line to which they belong.
Accordingly, at the next step, they take the appropriate colour. Note that as noted previously,
the yellow cell has the state~{\tt G}. The nine {\tt W}-cells turn to~{\tt A}.
Now, as can be checked on Figures~\ref{ndodprop} and~\ref{ndodZprop} as well as in
Table~\ref{dodrules}, faces~6 and~7 take the appropriate colour: they have the colour
of the cell they can see through their face~0 and through face~1 they can see another
{\tt A}-cell. The cell need not know which is its face~0. What is important is the number
of cells and the other non-blank colours. If there are two of them, if the yellow colour
is among them, remember that here the yellow colour is~{\tt G}, the cell turns to mauve, 
otherwise it turns to red. Accordingly, the process described in Subsection~\ref{continuation}
is correctly implemented.

    This allows us to transport the construction indicated by Figure~\ref{new_propa_penta}
in the case of the pentagrid. This can be seen
on Figures~\ref{ndodprop}, \ref{ndodHprop} and~\ref{ndodZprop}. Each figure consists
of sets of two pictures as mentioned in Figure~\ref{dodecviews}: one picture is the projection
onto~$\cal H$ from above of the cells in contact with this plane and which lie upon it. 
The other picture, somehow smaller, is the projection onto~$\cal H$ again, but from below, of 
the cells in contact with~$\cal H$ and which are below the plane. In the pictures of
Figures~\ref{ndodprop} and~\ref{ndodHprop} showing the projection above~$\cal H$, we can see 
the that the computation is exactly the same as in Figure~\ref{new_propa_penta}.

    Table~\ref{dodrules} also shows that the process illustrated by
Figure~\ref{ndodZprop} is correctly implemented. In particular, we note the rules
which make the last cells of the yellow and mauve lines disappear, involving the successive
vanishings of the green cells: when no more green cell is generated, the process stops
as the {\tt A}-cells with no support disappear.

\vtop{
\ligne{\hfill\includegraphics[scale=0.475]{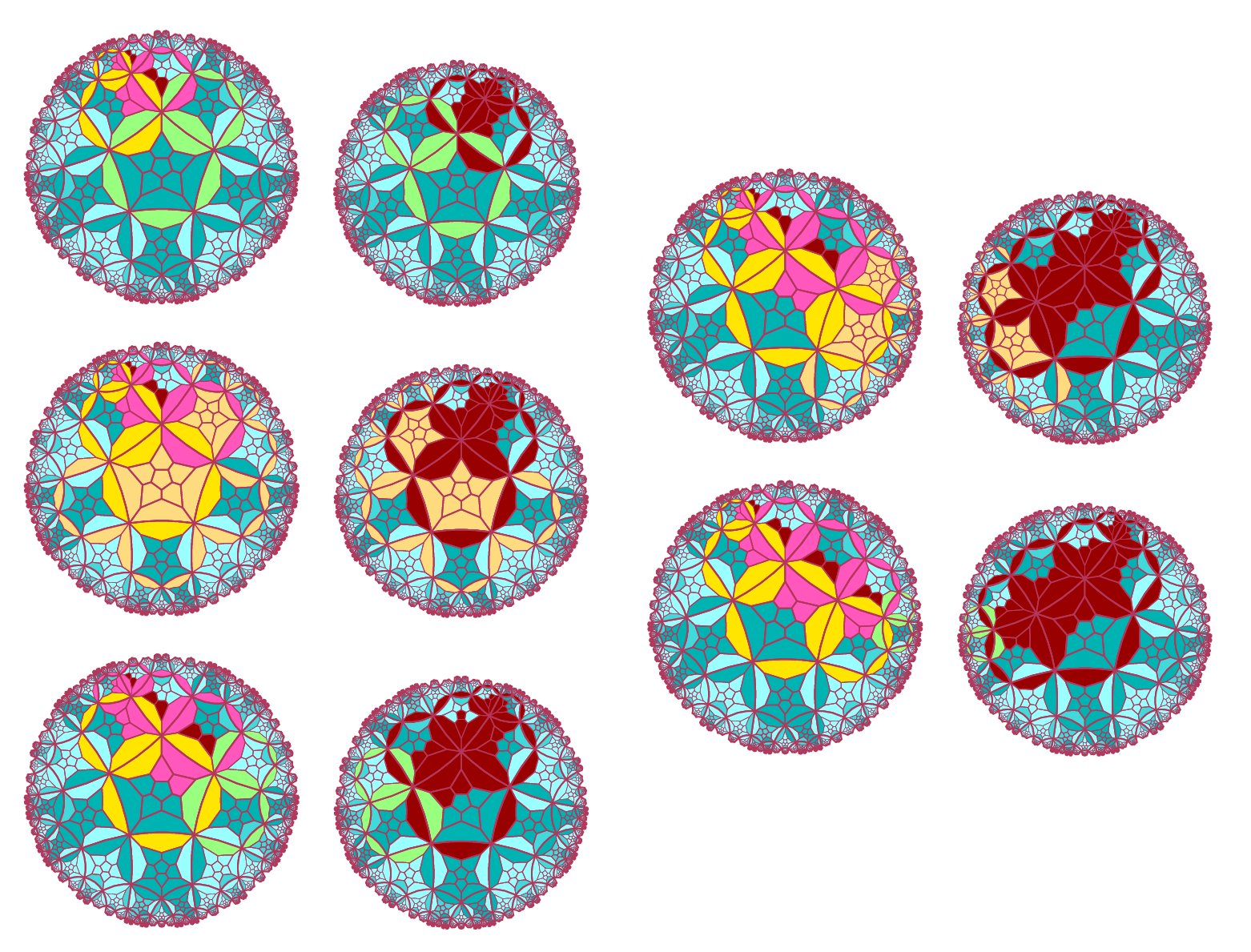}\hfill}
\vspace{-10pt}
\begin{fig}\label{ndodprop}
\leurre
The simultaneous construction of the lines.
\end{fig}
\vskip 8pt
}

\vtop{
\ligne{\hfill\includegraphics[scale=0.475]{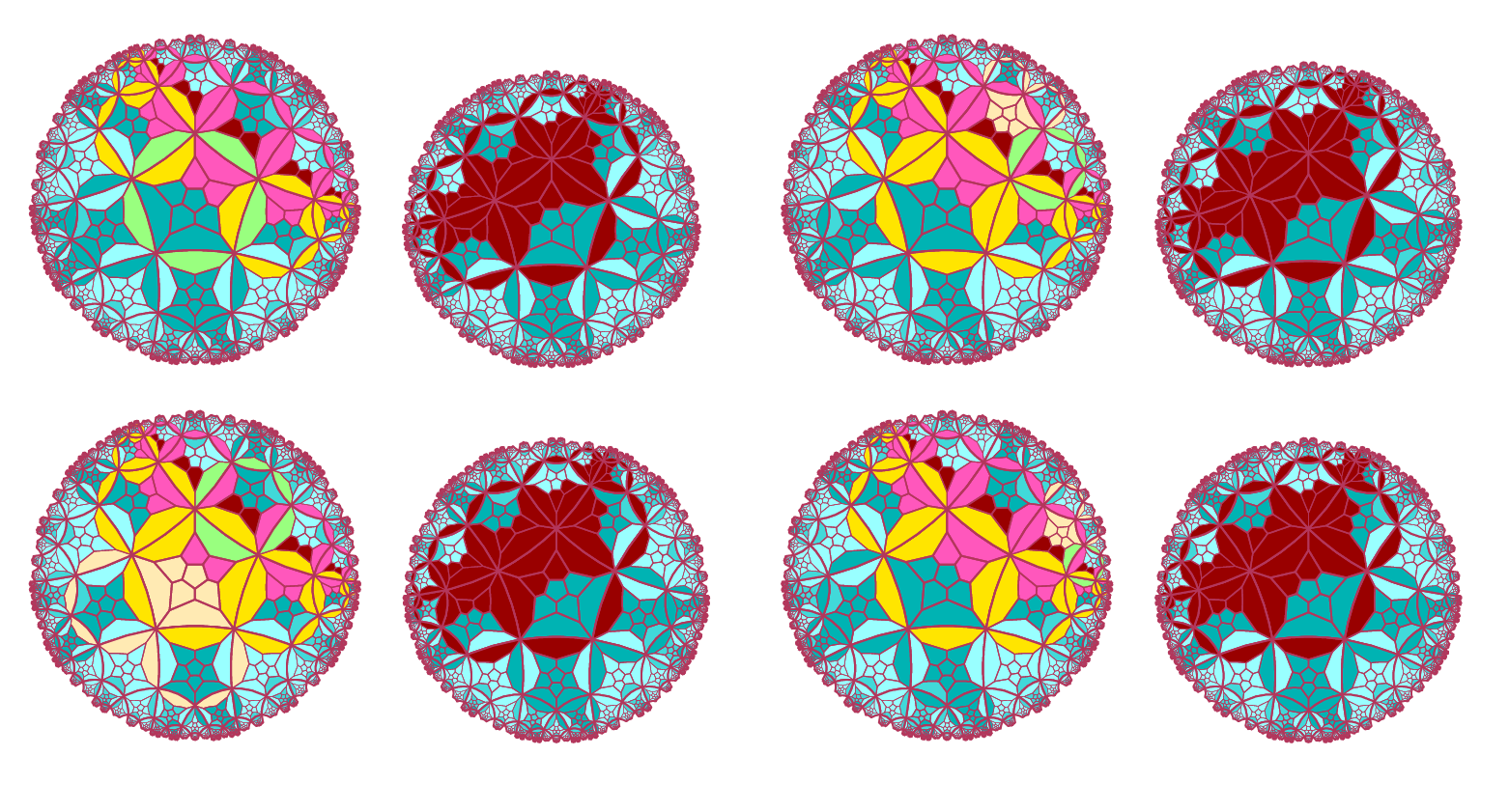}\hfill}
\begin{fig}\label{ndodHprop}
\vspace{-10pt}
\leurre
Propagation of the stopping signal along the mauve line. 
\end{fig}
}

\vtop{
\vspace{-10pt}
\ligne{\hfill\includegraphics[scale=0.475]{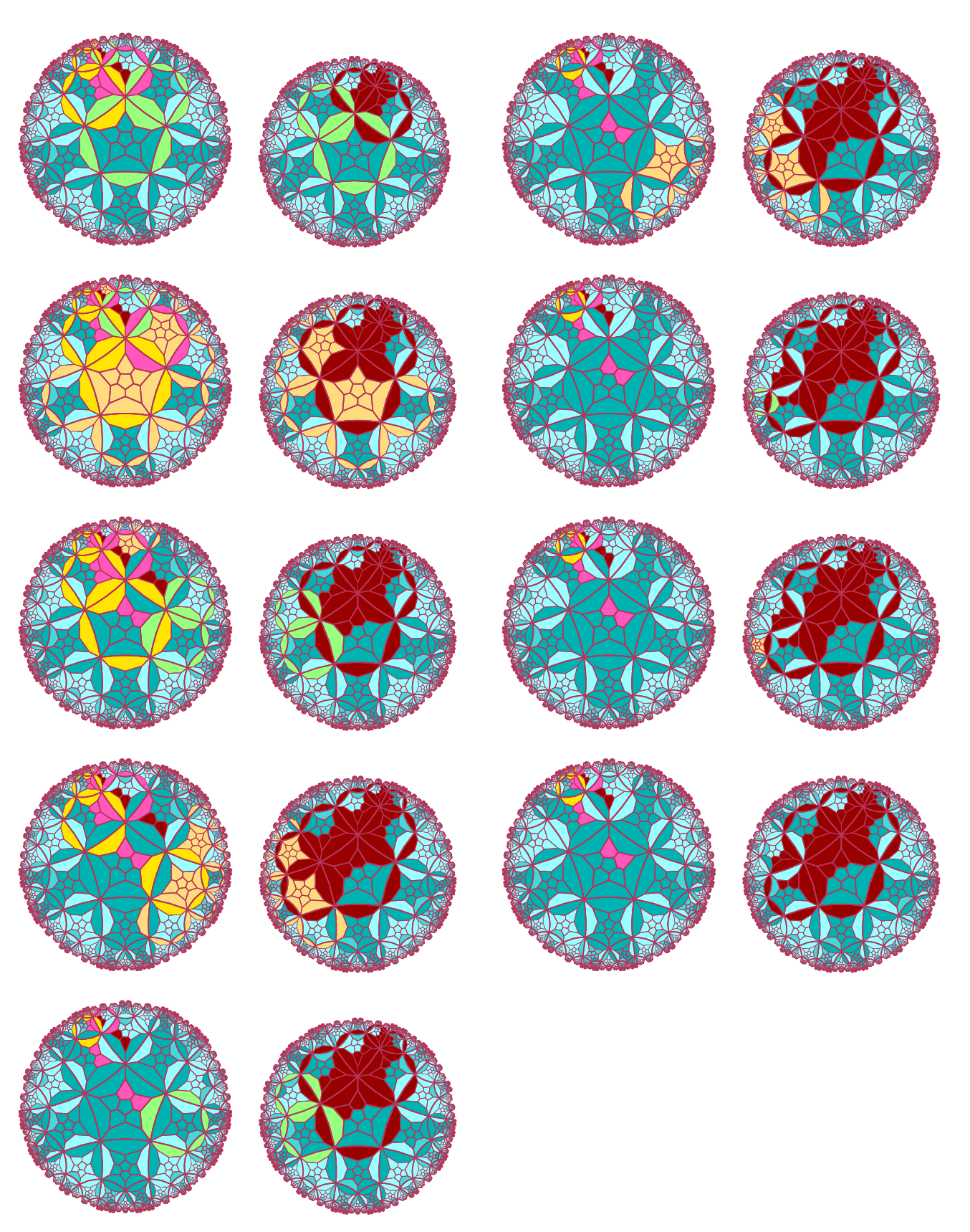}\hfill}
\begin{fig}\label{ndodZprop}
\vspace{-10pt}
\leurre
The stopping signal reaches the end and halts the computation.
\end{fig}
}

\subsubsection{Completing the proof of Theorem~\ref{stronghyp10}: the rules}
\label{rules10}

   In this subsection, we summarize the previous analyzes by giving the rules of the
cellular automaton which proves Theorem~\ref{stronghyp10}. We have an instance of the automaton
for each tiling we consider. We shall successively consider the case of the pentagrid, the
heptagrid and then the dodecagrid. In each case, we consider that the format of a rule
is \hbox{\tt w$_0$w$_1$w$_2$w$_3$w$_4$w$_5$w$_6$} in the pentagrid,
\hbox{\tt w$_0$w$_1$w$_2$w$_3$w$_4$w$_5$w$_6$w$_7$w$_8$} in the heptagrid
and \hbox{\tt w$_c$w$_0$w$_1$w$_2$w$_3$w$_4$w$_5$w$_6$w$_7$w$_8$w$_9$w$_{10}$w$_{11}$w$_n$}
for the dodecagrid where, in that latter case, {\tt w$_c$} is the current state of the
cell and~{\tt w$_n$} its new state after the application of the rule. In each case,
the {\bf context} of the rule~$\rho$ written as a word, is the subword~$\kappa$ such that
\hbox{$\rho = \kappa\hbox{\tt w}_n$} where {\tt w}$_n$ is the new state of the cell. 
  
\vskip 5pt
\ligne{\hskip-20pt$\vcenter{\underline{\hbox{Pentagrid}}}$
\hfill}
\vskip 5pt

The rules for the cellular automaton in this case are listed in 
Tables~\ref{new_rules_penta_yellow} and~\ref{new_rules_penta_red}. The first table gives the rule
for the cells belonging to the yellow line and their neighbours which do not belong to the
red one. The rules are adapted to the description of the initial configuration given in
Sub-subsection~\ref{ident}.

   Each table is divided into three parts: the continuation of the line, the occurrence of
the stopping signal and its progression on the red line to the end of the configuration and,
when the end is reached, the halting of the cellular automaton.

\vtop{
\begin{tab}\label{new_rules_penta_yellow}
\leurre
New rules for the pentagrid: continuation of the line, transfer of the stopping
signal and completion of the computation for the yellow cells and their
non-red neighbours. In the table, {\tt Y} replace any state of~\LN$_9$.
Different occurrences of~{\tt Y} in the same rule may represent different states.
Also, {\tt x} represents the symbols of the Turing machine: {\tt 0}, {\tt 1},
{\tt y} or~{\tt A}.
The rules are rotationally independent.
\end{tab}
\vspace{-12pt}
\grostrait
\setbox110=\vtop{\leftskip 0pt\parindent 0pt\hsize=130pt
\zz
\obeylines
\leftskip 0pt
\obeyspaces\global\let =\ \parskip=-2pt
--
--  yellow line and its 
--  non-red neighbours 
--
--  the computation
--  and continuing 
--  the line
--
--  0  1  2  3  4  5  6
--
    W  W  W  W  W  W  W 
    W  T  W  W  W  W  W 
    W  Y  W  W  W  W  W 
    B  T  T  B  B  W  B
    W  T  B  B  B  W  W
    W  T  W  B  B  W  B
    W  T  W  B  B  Y  W
    W  G  W  W  W  W  A
    W  G  B  W  W  W  A
\par
}
\setbox112=\vtop{\leftskip 0pt\parindent 0pt\hsize=130pt
\zz
\obeylines
\leftskip 0pt
\obeyspaces\global\let =\ \parskip=-2pt
--
--  0  1  2  3  4  5  6
--
    Y  T  Y  B  B  Y  Y
    A  U  W  W  W  W  W
    A  U  B  W  W  W  B
    A  A  U  W  W  W  G     
    G  G  U  W  W  W  U
    W  A  A  W  W  W  W
    B  Y  W  W  W  W  B
    B  Y  B  W  W  W  B  
    B  B  Y  W  W  W  B  
    U  T  U  A  A  A  U
--
--   stopping signal:
--   source and motion          
--
    T  T  0  B  B  y  G
    G  T  W  B  B  W  0
    W  T  G  B  B  W  W 
    W  T  W  B  B  G  W 
    0  G  W  B  B  W  0
\par
}
\setbox114=\vtop{\leftskip 0pt\parindent 0pt\hsize=130pt
\zz
\obeylines
\leftskip 0pt
\obeyspaces\global\let =\ \parskip=-2pt
--
--  0  1  2  3  4  5  6
--
    0  T  W  B  B  W  0
    x  G  W  B  B  W  x
    W  G  W  B  B  x  W
    W  G  x  B  B  W  W
    U  G  W  B  B  U  W
    B  A  W  W  W  W  B
--
--   halting of the 
--   computation        
--
    U  W  W  B  B  G  W
    U  W  W  A  A  A  W 
    W  W  W  B  B  G  W
    W  W  W  B  B  W  W
    G  W  W  W  W  W  W
    A  B  W  W  W  W  W
    A  T  A  W  W  W  G
\par
}
\ligne{\hfill\box110\hfill\box112\hfill\box114\hfill}
\vskip 5pt
\demitrait
}

\vtop{
\begin{tab}\label{new_rules_penta_red}
\leurre
New rules for the pentagrid: continuation of the line, transfer of the stopping
signal and completion of the computation for the red cells and their non-yellow neighbours.
The rules are rotationally independent.
\end{tab}
\vspace{-12pt}
\grostrait
\setbox110=\vtop{\leftskip 0pt\parindent 0pt\hsize=130pt
\zz
\obeylines
\leftskip 0pt
\obeyspaces\global\let =\ \parskip=-2pt
--
--  red line and its 
--  non-yellow neighbours 
--
--  the computation
--  and continuing 
--  the line
--
--  0  1  2  3  4  5  6
--
    T  B  T  W  W  W  T
    T  T  T  W  W  W  T
    T  T  Y  T  W  W  T
    T  T  U  A  A  A  T
    T  T  U  G  W  W  T
    G  T  G  W  W  W  T
\par
}
\setbox112=\vtop{\leftskip 0pt\parindent 0pt\hsize=130pt
\zz
\obeylines
\leftskip 0pt
\obeyspaces\global\let =\ \parskip=-2pt
--
--   stopping signal:
--   source and motion          
--
--  0  1  2  3  4  5  6
--
    T  T  G  T  W  W  G
    T  T  W  G  W  W  T
    T  G  Y  T  W  W  G
    W  A  G  W  W  W  W  
    T  T  0  G  A  A  T
    T  T  Y  G  A  W  T
    G  T  Y  T  W  W  T 
\par
}
\setbox114=\vtop{\leftskip 0pt\parindent 0pt\hsize=130pt
\zz
\obeylines
\leftskip 0pt
\obeyspaces\global\let =\ \parskip=-2pt
--
--   halting of the 
--   computation        
--
--  0  1  2  3  4  5  6
--
    T  G  U  G  W  W  G
    T  G  U  A  A  A  W
    T  T  W  W  W  W  T
    W  T  U  G  W  W  W
    G  G  W  W  W  W  W
    W  W  U  A  A  A  W
    A  A  W  W  W  W  W
\par
}
\ligne{\hfill\box110\hfill\box112\hfill\box114\hfill}
\vskip 5pt
\demitrait
\vskip 8pt
}

   The rules were established from a careful analysis of Figure~\ref{new_propa_penta},
going cell after cell along the lines, first the yellow one and then the red one.
   It is not difficult to check that there is no contradiction in the rules and that they
completely describe the working of the cellular automaton. This completes the proof 
of Theorem~\ref{stronghyp10} for the pentagrid. They are also rotation invariant.
  
\vskip 5pt
\ligne{\hskip-20pt$\vcenter{\underline{\hbox{Heptagrid}}}$
\hfill}
\vskip 5pt

   The rules for the cellular automaton in the heptagrid are listed in 
Table~\ref{new_rules_hepta_whole}. They are also established from Figure~\ref{new_propa_hepta}
exactly in the same way as the rules for the pentagrid were deduced from 
Figure~\ref{new_propa_penta}.
As can be seen from Figure~\ref{new_propa_hepta} there are much less rules than in the case 
of the heptagrid.
This is why we have a single table. The rules for the cells of the yellow line need two
columns while the last one is devoted to the cells of the red line. Here too, we can check that
the rules are coherent, rotation invariant and they completely describe the motion of the
cellular automaton.

\vtop{
\begin{tab}\label{new_rules_hepta_whole}
\leurre
New rules for the heptagrid: continuation of the line, transfer of the stopping
signal and completion of the computation. First two columns: the rules for the yellow 
cells and their non-red neighbours. Third column: the rules for the red cells and their
non-yellow neighbours. In the table, {\tt Y} replace any state of~\LN$_9$.
Different occurrences of~{\tt Y} in the same rule may represent different states.
Also, {\tt x} represents the symbols of the Turing machine: {\tt 0}, {\tt 1},
{\tt y} or~{\tt A}.
The rules are rotationally independent.
\end{tab}
\vspace{-12pt}
\grostrait
\setbox110=\vtop{\leftskip 0pt\parindent 0pt\hsize=130pt
\zz
\obeylines
\leftskip 0pt
\obeyspaces\global\let =\ \parskip=-2pt
--
--  yellow line and its 
--  non-red neighbours 
--
--  the computation
--  and continuing 
--  the line
--
--  0  1  2  3  4  5  6  7  8
--
    W  W  W  W  W  W  W  W  W
    W  T  W  W  W  W  W  W  W
    W  Y  W  W  W  W  W  W  W
    B  B  W  T  T  T  W  W  B
    W  T  B  B  W  B  W  T  W
    Y  T  Y  B  W  B  Y  T  Y
    B  Y  Y  W  W  W  W  W  B
    B  G  U  W  W  W  W  W  B
    W  G  B  W  W  W  W  W  W
    W  T  G  W  W  W  W  W  A
    W  A  G  W  W  W  W  W  G
    W  G  B  W  W  W  W  W  W
    W  B  Y  B  W  W  W  W  W
    G  T  U  W  W  W  W  W  G
    G  A  T  U  B  W  W  W  U
    W  G  U  W  W  W  W  W  B
\par
}
\setbox112=\vtop{\leftskip 0pt\parindent 0pt\hsize=130pt
\zz
\obeylines
\leftskip 0pt
\obeyspaces\global\let =\ \parskip=-2pt
--
--   stopping signal:
--   source and motion          
--
--  0  1  2  3  4  5  6  7  8
--
    T  T  0  B  W  B  y  T  G
    0  T  W  B  W  B  T  T  W
    y  T  T  B  W  B  W  T  W
    B  G  W  W  W  W  W  W  B
    Y  T  Y  B  W  B  Y  G  Y
    Y  G  Y  B  W  B  Y  T  Y
    U  T  Y  B  W  B  U  G  W
    W  T  Y  B  W  B  U  G  W
--
--   halting of the 
--   computation        
--
    U  G  U  B  W  B  G  T  W
    W  T  W  B  W  B  U  W  W
    W  T  W  B  W  B  U  W  W
    U  T  W  W  B  W  W  G  W
    B  U  W  W  W  W  W  W  B
    B  G  W  W  W  W  W  W  B
    B  W  W  W  W  W  W  W  B
    W  T  U  B  W  B  U  W  W
    W  T  U  B  W  B  W  W  W
    A  G  W  W  W  W  W  W  W
    G  A  W  W  B  W  W  W  W
\par
}
\setbox116=\vtop{\leftskip 0pt\parindent 0pt\hsize=130pt
\zz
\obeylines
\leftskip 0pt
\obeyspaces\global\let =\ \parskip=-2pt
--
--  red line and its 
--  non-yellow neighbours 
--
--  the computation
--  and continuing 
--  the line
--
--  0  1  2  3  4  5  6  7  8
--
    T  B  T  W  W  W  W  W  T
    T  T  B  T  W  W  W  W  T
    T  T  U  G  W  W  W  W  T
--
--   stopping signal:
--   source and motion          
--
    T  T  W  G  T  W  W  W  G
    T  T  G  W  T  W  W  W  T
    T  T  Y  Y  G  W  W  W  T
    T  G  Y  Y  T  W  W  W  G
    G  T  Y  Y  T  W  W  W  T
--
--   halting of the 
--   computation        
--
    T  G  U  G  A  W  W  W  W
    T  U  G  W  W  W  W  W  W
\par
}
\ligne{\hfill\box110\hfill\box112\hfill\box116\hfill}
\vskip 5pt
\demitrait
\vskip 10pt
}

This completes the proof of Theorem~\ref{stronghyp10} in the case of the heptagrid. 

\vskip 5pt
\ligne{\hskip-20pt$\vcenter{\underline{\hbox{Dodecagrid}}}$
\hfill}
\vskip 5pt

   Table~\ref{dodrules} gives the rules for the computation of the cellular automaton in the
case of the dodecagrid. The rules were established from an analysis of 
Figures~\ref{ndodprop}, \ref{ndodHprop} and~\ref{ndodZprop}.

    We remark that the scenario is very close to what was performed in the pentagrid.
As in the previous tables, the states are those of~\LN$_9$ to which we append the
state~{\tt G}. As noted in the study of the figures, most of the cells have a structure
in which six faces are blank forming a ring around a face in a single way. This remarkable 
feature allows us to establish the rotation invariance of the cellular automaton as well
as to check the coherence of the rules. Also, the rules completely describe the
working of the automaton.

This completes the proof of Theorem~\ref{stronghyp10} in the case of the dodecagrid. 

\section{Conclusion}

   Theorems~\ref{stronghyp11} and~\ref{stronghyp10} significantly improve the result 
of~\cite{mmstrong,mmbook3}. As far as known to the author, Theorem~\ref{strong1D} is
the best result on a small strongly universal cellular automaton on the line.

\setbox110=\vtop{\leftskip 0pt\parindent 0pt\hsize=240pt
\zz
\obeylines
\leftskip 0pt
\obeyspaces\global\let =\ \parskip=-2pt
--
 c  0  1  2  3  4  5  6  7  8  9 10 11  n
--
 W  W  W  W  W  W  W  W  W  W  W  W  W  W
 W  B  W  W  W  W  W  W  W  W  W  W  W  W
 W  Y  W  W  W  W  W  W  W  W  W  W  W  W
 W  T  W  W  W  W  W  W  W  W  W  W  W  W
--
--  start and computation
--
 B  T  B  W  W  W  U  T  T  W  W  W  W  B
 B  T  B  W  W  W  W  B  B  W  W  W  W  B
 W  T  B  W  W  W  B  B  B  W  W  W  W  W
 Y  T  B  Y  W  W  Y  B  B  W  W  W  W  Y
 T  B  T  T  W  W  W  T  T  W  W  W  W  T 
 T  Y  T  T  W  W  T  T  T  W  W  W  W  T 
 U  T  B  G  W  W  U  B  B  W  W  W  W  U
 B  T  Y  B  W  W  G  T  T  W  W  W  W  B
 T  B  T  G  W  W  T  T  T  W  W  W  W  T 
 T  Y  T  T  W  W  G  T  T  W  W  W  W  T 
 W  G  W  W  W  W  W  W  W  W  W  W  W  A
 G  G  G  W  W  W  U  W  W  W  W  W  W  U 
 G  G  G  B  W  W  W  W  W  W  W  W  W  B
 G  G  G  T  W  W  W  W  W  W  W  W  W  T
 G  G  G  W  W  W  T  W  W  W  W  W  W  T
--
-- continuing the line
--
 B  T  Y  B  W  W  B  T  T  W  W  W  W  B
 Y  T  B  Y  W  W  Y  B  B  W  W  W  W  Y
 B  T  Y  B  A  A  A  A  A  A  A  A  A  B
 Y  T  B  A  A  A  Y  A  A  A  A  A  A  Y
 T  B  T  T  W  W  T  T  T  W  W  W  W  T    
 T  Y  T  T  W  W  T  T  T  W  W  W  W  T    
 T  B  T  T  A  A  A  A  A  A  A  A  A  T  
 T  Y  T  A  A  A  T  A  A  A  A  A  A  T  
 A  B  W  W  W  W  W  W  W  W  W  W  W  W
 A  Y  W  W  W  W  W  W  W  W  W  W  W  W
 A  T  W  W  W  W  W  W  W  W  W  W  W  W
 A  B  T  A  W  W  W  W  W  W  W  W  W  T 
 A  B  A  W  W  W  W  W  W  W  W  W  W  T
\par
}
\setbox112=\vtop{\leftskip 0pt\parindent 0pt\hsize=240pt
\zz
\obeylines
\leftskip 0pt
\obeyspaces\global\let =\ \parskip=-2pt
--
 c  0  1  2  3  4  5  6  7  8  9 10 11  n
--
 A  Y  A  B  W  W  W  W  W  W  W  W  W  B
 A  Y  A  W  W  W  W  W  W  W  W  W  W  B
 A  T  A  B  W  W  W  W  W  W  W  W  W  T 
 A  T  A  W  W  W  W  W  W  W  W  W  W  T 
 A  T  B  A  W  W  W  W  W  W  W  W  W  T
 A  A  Y  A  W  W  W  W  W  W  W  W  W  G
 A  A  A  B  W  W  W  W  W  W  W  W  W  G
 A  A  A  T  W  W  W  W  W  W  W  W  W  G
 A  A  T  A  W  W  W  W  W  W  W  W  W  G
 W  A  A  W  W  W  W  W  W  W  W  W  W  W
 W  B  B  W  W  W  W  W  W  W  W  W  W  W
 W  T  T  W  W  W  W  W  W  W  W  W  W  W
--
-- stopping signal
--
 T  T  B  y  W  W  0  B  B  W  W  W  W  G
 y  T  B  W  W  W  T  B  B  W  W  W  W  Y
 0  T  B  T  W  W  W  B  B  W  W  W  W  Y
 G  T  B  W  W  W  W  B  B  W  W  W  W  0
 W  T  B  W  W  W  G  B  B  W  W  W  W  W 
 W  T  B  G  W  W  W  B  B  W  W  W  W  W 
 B  T  G  B  W  W  B  T  T  W  W  W  W  G
 G  T  0  B  W  W  B  T  T  W  W  W  W  B
 B  T  Y  B  W  W  G  T  T  W  W  W  W  B 
 B  T  Y  G  W  W  B  T  T  W  W  W  W  G 
 x  T  G  W  W  W  W  B  B  W  W  W  W  x
 U  T  G  U  W  W  W  B  B  W  W  W  W  W
--
-- halting the computation
--
 B  T  U  G  A  A  A  A  A  A  A  A  A  W
 G  G  G  W  W  W  W  W  W  W  W  W  W  W
 U  T  W  W  W  W  G  B  B  W  W  W  W  W
 U  T  U  W  W  W  W  B  B  W  W  W  W  W
 U  T  W  A  A  A  W  A  A  A  A  A  A  W
 W  T  W  W  W  W  W  B  B  W  W  W  W  W
 G  G  W  W  W  W  W  W  W  W  W  W  W  W
 G  W  G  W  W  W  T  W  W  W  W  W  W  W
 G  W  W  T  W  W  W  W  W  W  W  W  W  W
\par}
\vtop{
\begin{tab}\label{dodrules}
\leurre
Rules for the propagation of the $1D$-structure in the dodecagrid and for the stopping
of the computation of~$\cal P$ once \LN{} detected the halting of the simulated computation.
Remember that the column~{\tt c} represents the current state of a cell of the automaton
and that the column~{\tt n} represents its new state after the application of the rule.
The rules are rotationally independent.
\end{tab}
\vspace{-12pt}
\grostrait
\ligne{\hfill\box110\hfill\box112\hfill}
\vskip 7pt
\demitrait
\vskip 7pt
}

\subsection*{Acknowledgement}

   The author is much in debt for the reviewers for their remarks. He is much in debt to 
Turlough Neary for the discussions mentioned in Subsection~\ref{exec}. He is also much in 
debt to the organizers of {\bf MCU'2013} for accepting the paper.

\nocite{*}
\bibliographystyle{eptcs}
\bibliography{mm_new_deffinal}

\end{document}